\PassOptionsToPackage{dvipsnames}{xcolor}

\documentclass[traditabstract]{aa}
\usepackage{verbatim} 
\usepackage{amsmath}
\usepackage{amssymb}
\usepackage{graphicx}
\usepackage{threeparttable}
\usepackage{rotating} 
\usepackage{booktabs}
\usepackage[authoryear]{natbib}
\usepackage[colorlinks=true,linkcolor=blue,urlcolor=blue,citecolor=blue]{hyperref}
\usepackage{xspace}
\usepackage{siunitx}
\usepackage{subcaption}  

\usepackage{tikz}
\usepackage{adjustbox} 
\usetikzlibrary{calc}  
\usepackage{xcolor}
\makeatletter

\usepackage{txfonts}\usepackage{twoopt}
\bibpunct{(}{)}{;}{a}{}{,}
\usepackage{enumitem}

\newcommand{\eg}{e.g.\@\xspace}
\newcommand{\cf}{c.f.\@\xspace}

\titlerunning{Interaction-powered light curves from binary-driven CSM}
\authorrunning{Dandan Wei et al.}

\makeatother

\begin{document}

\title{Diversity of stripped-envelope supernova light curves from interaction with binary-driven circumstellar material}

\author{Dandan Wei\inst{\ref{ISTA}}\thanks{dandan.wei@hotmail.com},
Fabian R. N. Schneider\inst{\ref{HITS},\ref{ZAH}},    
Philipp Podsiadlowski\inst{\ref{Oxford}},
Takashi J. Moriya\inst{\ref{NAOJ},\ref{JapanGIAS},\ref{Monash}},
Eva Laplace\inst{\ref{HITS},\ref{Leuven1},\ref{Leuven2},\ref{Amsterdam}}
}

\institute{Institute of Science and Technology Austria, 
Am Campus 1, 3400 Klosterneuburg, Austria\label{ISTA}
\and Heidelberger Institut f\"{u}r Theoretische Studien, Schloss-Wolfsbrunnenweg 35, 69118 Heidelberg, Germany\label{HITS}
\and Zentrum f\"{u}r Astronomie der Universit\"{a}t Heidelberg, Astronomisches Rechen-Institut, M\"{o}nchhofstr. 12-14, 69120 Heidelberg, Germany\label{ZAH}
\and Department of Physics, University of Oxford, Denys Wilkinson Building, Keble Road, Oxford OX1 3RH, United Kingdom\label{Oxford}
\and National Astronomical Observatory of Japan, National Institutes of Natural Sciences, 2-21-1 Osawa, Mitaka, Tokyo 181-8588, Japan \label{NAOJ}
\and Graduate Institute for Advanced Studies, SOKENDAI, 2-21-1 Osawa, Mitaka, Tokyo 181-8588, Japan \label{JapanGIAS}
\and School of Physics and Astronomy, Monash University, Clayton, VIC 3800, Australia \label{Monash}
\and Institute of Astronomy, KU Leuven, Celestijnenlaan 200D, B-3001 Leuven, Belgium \label{Leuven1}
\and Leuven Gravity Institute, KU Leuven, Celestijnenlaan 200D, Box 2415, 3001 Leuven, Belgium \label{Leuven2}
\and Anton Pannekoek Institute of Astronomy, University of Amsterdam, Science Park 904, 1098 XH Amsterdam, The Netherlands \label{Amsterdam}
}

\date{Received xxx / Accepted yyy}

\abstract{
A growing number of core-collapse supernovae (CCSNe) exhibit diverse light-curve morphologies that indicate strong interaction with dense, pre-existing circumstellar material (CSM). Understanding the physical origin of such CSM is essential for exploring the late-stage evolution of SN progenitors and the diversity of observed transients. Non-conservative mass transfer during binary interactions provides a promising channel for producing dense CSM before core collapse. Using the stellar evolution code MESA, we simulate post-common-envelope binaries that give rise to ultra-stripped progenitors before core collapse, and self-consistently construct the CSM from the mass loss during binary evolution. We further analytically predict the resulting bolometric and radio light curves by following the shock dynamics of SN ejecta interacting with the CSM. We find that surface-radius variations in the ultra-stripped progenitors, which reflect changes in core energy generation after a delay comparable to the thermal timescale, trigger multiple episodes of mass transfer and give rise to a variety of CSM density profiles, including detached shells and multi-peaked structures. When interacting with SN ejecta, such CSMs give rise to non-monotonic, multi-peaked optical and radio light curves that qualitatively resemble features observed in some stripped-envelope SNe. Future long-term multi-wavelength monitoring of ultra-stripped SN candidates, particularly at late times, will be crucial for probing structured CSM and constraining the long-term mass-loss history of stripped progenitors prior to core collapse. Our results suggest that SN-CSM interaction in binaries hosting ultra-stripped progenitors can provide a possible channel for producing diverse light-curve morphologies, highlighting a potential connection between the pre-SN evolution of massive binary stars and the diversity of their explosive transients.
}


\maketitle

\section{Introduction}
\label{sec:introduction}
With the rapid development of transient surveys, a significant fraction of core-collapse supernovae (CCSNe) progenitors are inferred to be surrounded by dense circumstellar materials (CSM) \citep{2020ApJPerley,2021ApJBruch}. When the SN ejecta collide with the pre-existing slow-moving CSM, observational signatures of ejecta-CSM interaction emerge. Hydrogen-rich SNe exhibiting narrow emission lines are classified as Type IIn SNe \citep{1990MNRASSchlegel}, while their hydrogen-poor counterparts are identified as Type Ibn and Type Icn SNe \citep{2008MNRASPastorello,2019ApJPooley,2022ApJPellegrino}. These interacting SNe display diverse light-curve structures owing to the additional energy input powered by the ejecta-CSM interaction. Some of them show rapidly rising and luminous early-time light curves, observed not only in hydrogen-rich events \citep{2017NatPhYaron,2018NatAsForster,2020MNRASZhang,2021ApJBruch,2024ApJJacobson}, but also in hydrogen-poor SNe (referred to as stripped-envelope CCSNe) \citep{2020MNRASClark,2019ApJHo,2020ApJYao,2020ApJGangopadhyay,2022NaturGalYam}. In addition to early-time interaction signatures, several SNe also exhibit late-phase evidence of CSM interaction, most notably late-time radio re-brightening \citep{2021ApJStroh,2024MNRASRose}, as observed in SN 2004dk \citep{2018MNRASMauerhan, 2021ApJBalasubramanian}, SN 2014C \citep{2017ApJMargutti}, and SN 2017ens \citep{2018ApJChen}. The presence of such CSM around interacting SN progenitors provides a crucial bridge for better understanding the last-stage evolution of massive stars and the explosion mechanisms that give rise to diverse transients.

Increasing observational evidence suggests that the dense CSM surrounding the interacting SN progenitors originates from substantial mass-loss episodes that occur shortly before core collapse. However, the physical mechanisms driving such mass loss remain uncertain. From the perspective of a single-stellar evolution, several processes have been proposed, including wave-driven mass loss triggered by super-Eddington core luminosities \citep{2012MNRASQuataert,2014ApJShiode,2014ApJSmith,2017MNRASFuller,2020ApJMorozova,2021ApJWu}, pulsations of red supergiant SN progenitors \citep{2010ApJYoon,2015A&AMoriya,2025A&ABronner,2026ApJLaplace}, and unstable silicon burning leading to violent flashes in the stellar core \citep{2015ApJWoosley}. In addition, binary interactions in massive binary systems have recently been proposed as an efficient channel to enhance the mass loss before explosion \citep{2012ApJChevalier, 2022ApJWu,2024ApJMatsuoka,2024A&AErcolino,2025A&AErcolino,2024ApJDong,2025ApJWu}. 

Massive stars that lose most of their hydrogen-rich envelopes through binary interaction can end their lives as stripped-envelope SNe. During the late evolutionary stages -- particularly after central helium burning -- the radii of these stripped SN progenitors may expand again \citep{2020A&ALaplace}, triggering additional episodes of mass transfer before core collapse. Material can be expelled from the binary system during non-conservative mass transfer, potentially forming the CSM around stripped-envelope SN progenitors \citep{2022ApJWu,Wei2024A&A,2025ApJWu,2025A&AErcolino}. In close binaries, such additional stripping can lead to ultra-stripped progenitors, first modelled in detailed binary evolution calculations by \citet{2013ApJTauris, 2015MNRASTauris}, where the helium-rich envelope can be almost completely stripped through the additional mass transfer to a compact companion. CSM with detached shells can form through the mass transfer from low-mass stripped stars (with mass less than $3\,M_{\odot}$) due to their non-monotonic radius evolution, whereas more massive stripped stars tend to produce continuous, wind-like CSM profiles \citep{2024ApJDong,2025ApJWu}. 

Radio observations provide a powerful probe for exploring the CSM surrounding SN progenitors. Approximately $95\%$ of all SN remnants identified in the Milky Way have been detected at radio wavelengths \citep{2014BASIGreen}, and the radio emission observed in many stripped-envelope SNe indicates the presence of dense CSM \citep{2021ApJMaeda, 2021ApJStroh, 2024MNRASRose}. Interaction with dense, detached CSM shells has been proposed to explain the late-time radio re-brightening \citep{2021ApJStroh, 2025ApJWu}. However, it remains challenging for current binary-evolution models involving non-conservative mass transfer to reproduce the non-monotonic light curves, especially the diverse early-time radio observation \citep{2025ApJWu}. In this paper, we present that material expelled during the non-conservative mass transfer in our post-common-envelope binary models can produce non-monotonic radio light curves that qualitatively resemble some observed features of stripped-envelope SNe, with radio luminosities comparable to those observed in some stripped-envelope SNe.

In this study, we self-consistently investigate how interaction-powered light curves may arise from the dense, diverse CSM formed through late-stage binary mass loss of ultra-stripped progenitors. The CSM density profiles are constructed based on the binary mass-loss history and the dynamical evolution of outflows launched through the outer Lagrange point. The bolometric and radio light curves are predicted by following the shock dynamics resulting from the collision between SN ejecta and the dense CSM. We find that the time-variable mass-loss rates during the non-conservative mass transfer result in non-monotonic features on the CSM, which in turn produce a wide range of interaction-powered light-curve morphologies. Our results suggest that interaction-powered light curves may provide a valuable probe of the late-stage evolution of SN progenitors and the binary interactions that shape their final mass-loss histories.

This paper is organised as follows. In Sect.\,\ref{sec:method}, we outline our methodology, including the binary evolution models adopted in this work, the construction of the CSM density profiles, and the analytical framework used to compute the interaction-powered bolometric and radio light curves. In Sect.\,\ref{results part}, we present the resulting CSM structures and the corresponding SN light curves, focusing on both the bolometric light curves and the radio emission, and compare these with several stripped-envelope SN observations. We discuss our findings in Sect.\,\ref{Discussion} and summarise the main conclusions in Sect.\,\ref{Conculsion}.

\section{Method}
\label{sec:method}
In this section, we consider non-conservative mass transfer in post-common-envelope binaries (Sect.\,\ref{sec:method-binary models}) giving rise to pre-SN CSM structures (Sect.\,\ref{method:CSM}). We model the shock evolution arising from the interaction between the SN ejecta and the CSM, and compute the corresponding bolometric light curves (Sect.\,\ref{sec: method interaction-powered luminosity}) as well as the radio light curves produced by synchrotron emission (Sect.\,\ref{method: radio emission}).

\subsection{Binary evolution models}
\label{sec:method-binary models}

Binary models with the same initial configurations as the post-common-envelope (post-CE) massive binary models presented in \cite{Wei2024A&A}, are simulated using MESA r12778 \citep{2011ApJSPaxton, 2013ApJSPaxton, 2015ApJSPaxton, 2018ApJSPaxton, 2019ApJSPaxton, 2023ApJSJermyn}. These binary systems are the remnants of 3-dimensional magneto-hydrodynamic CE simulations. Each system contains an envelope-stripped helium star in a short-period orbit around a compact companion -- either a neutron star or a black hole (treated as point masses). To obtain a more detailed picture of the mass-loss history before core collapse, we evolve all models to core silicon depletion, defined as the stage at which the central silicon mass fraction drops below $10^{-3}$.

The binary modelling generally follows the approach in \cite{Wei2024A&A}, and we refer the reader to this paper for details. For binary mass transfer, the ``Kolb'' mass-transfer scheme is applied \citep{1990A&AKolb}, with $50\%$ of the transferred material accreted by the companions. Accretion onto the compact companion during the binary interaction is limited to the corresponding Eddington-limited accretion rate, namely $1.8 \times 10^{-8}\,\rm M_{\odot}\,\rm yr^{-1}$ for the NS companion and $1.9 \times 10^{-7}\,\rm M_{\odot}\,\rm yr^{-1}$ for the BH companion. Our work includes two improvements that enhance our ability to characterise the formation of the CSM. First, the wind mass loss of the stripped helium star is calculated throughout its entire evolution using MESA’s ``Dutch'' wind prescription \citep{2009A&AGlebbeek}. In this scheme, the mass-loss of \citet{1988A&ASdeJager} is adopted for cool stars with eﬀective temperature below 10 000 K, while for hotter stars the prescription of \citet{2001A&AVink} is applied, when the surface hydrogen mass fraction exceeds 0.4, switching to \citet{2000A&ANugis} when it is below 0.4. Second, the upper limit of the mass-transfer rate in the binary modelling is increased from $10^{-3}\,\rm M_{\odot}\, yr^{-1}$ to $1.0\,\rm M_{\odot}\, yr^{-1}$, allowing for rapid late-stage mass transfer episodes to be captured better. These modifications have negligible effects on the overall binary evolution and its final fate, but they can affect the detailed mass-loss history and therefore the CSM density structure before core collapse. This, in turn, influences the observable signature of the subsequent ejecta-CSM interaction.

\subsection{Density structure of CSM}
\label{method:CSM}

\begin{figure*}
\begin{center}
\includegraphics[width=\textwidth]{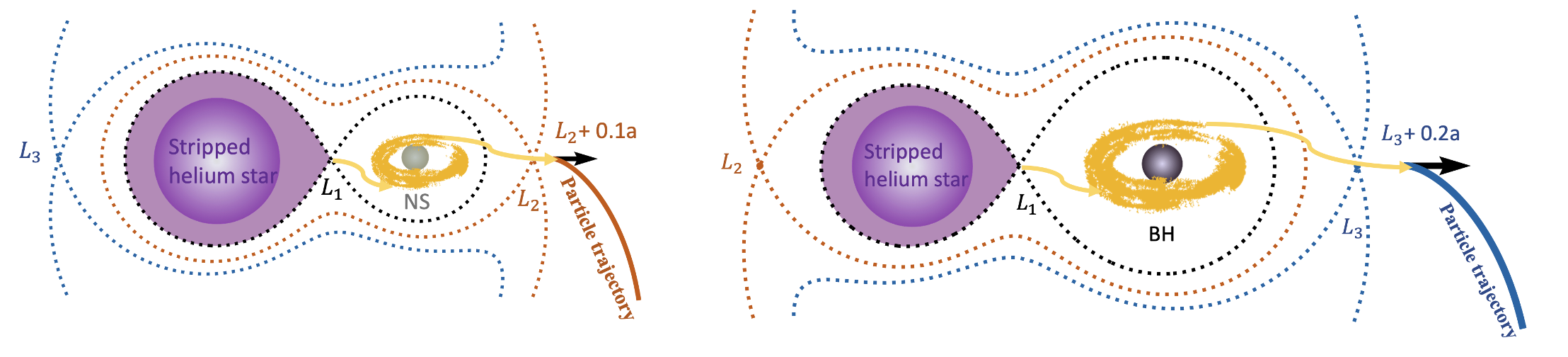}
\caption{Illustration of outflows from outer Lagrange points ($L_{\rm2}$ or $L_{\rm3}$). The dotted lines are equipotentials for the Roche potential in the equatorial plane of the binary system. In the left panel, the unaccreted material escapes from a position $0.1a$ beyond $L_{\rm2}$ in the case of a NS companion, while in the right panel, material leaves from a position $0.2a$ beyond $L_{\rm3}$ for a BH companion. Here $a$ is the separation of the binary system.}
\label{fig: trajectory cartoons}
   \end{center}
\end{figure*}

In our binary models, material is expelled from the system through two primary channels: the outflow from the vicinity of the outer Lagrangian point during mass transfer (hereafter referred to as RLOF outflow) and the stellar wind from the stripped helium star. These two mass-loss mechanisms together build up the CSM surrounding the SN progenitor. Owing to their different mass-loss rates and characteristic velocities, the resulting CSM via two channels exhibit distinct density profiles.

The total, time-dependent CSM density is obtained by combining the contributions from the stellar wind and the RLOF outflow,
\begin{equation}
\label{eqs.1}
\rho_{\rm csm} = \rho_{\rm wind} + \rho_{\rm RLOF}. 
\end{equation}
The detailed calculations of $\rho_{\rm wind}$ and $\rho_{\rm RLOF}$ are described in Sect.\,\ref{sec: method wind mass loss} and Sect.\,\ref{sec: method RLOF mass loss}, respectively.

In this work, we focus on the CSM density profile at the moment of core collapse, since this is the CSM structure that subsequently interacts with the rapidly expanding SN ejecta.

\subsubsection{Stellar wind}
\label{sec: method wind mass loss}

Material lost via stellar winds is assumed to be ejected isotropically. The wind density at radius $R_{\rm wind}$ is therefore given by
\begin{equation}
\label{eqs.2}
\rho_{\rm wind} = \frac{\dot{M}_{\rm wind}}{4 \pi {R^2_{\rm wind}} \upsilon_{\rm wind}}
\end{equation}
where $\dot{M}_{\rm wind}$ is the wind mass-loss rate from the stripped helium star, and $\upsilon_{\rm wind}$ is taken to be its surface escape velocity,
\begin{equation}
\label{eqs.3}
\upsilon_{\rm wind} = \sqrt{\frac{2GM_{1}}{R_{1}}}
\end{equation}
with $M_{1}$ and $R_{1}$ denoting its mass and radius, respectively.

The radial distance reached by the wind material by the time of core collapse can be estimated as  
\begin{equation}
\label{eqs.4}
R_{\rm wind} = \upsilon_{\rm wind}\,\tau_{\rm rem},
\end{equation}
where $\tau_{\rm rem}$ is the remaining lifetime of the stripped helium star until the end of our computation.

\subsubsection{RLOF outflow}
\label{sec: method RLOF mass loss}
During binary interactions, most of the transferred material cannot be accreted by the compact companion because of Eddington-limited accretion (see Sect.\,\ref{sec:method-binary models}). We assume that this non-accreted material escapes the binary system through the vicinity of the outer Lagrangian point, forming an outflow referred to as RLOF outflow (see Fig.\,\ref{fig: trajectory cartoons}). 

In a reference frame corotating with the binary system, the Roche potential at position $\vec{r}$ is given by
\begin{equation}
\label{eqs.5}
\psi \left( \vec{r} \right) = - G\,\frac{m_1}{\left|\,\vec{r}_1 - \vec{r}\,\right|} - G\,\frac{m_2}{\left|\,\vec{r}_2 - \vec{r}\,\right|} - \frac{1}{2} \left| \, \Omega \times \vec{r} \, \right|^2,
\end{equation}  
where $\vec{r}_1$ and $\vec{r}_2$ are the position vectors relative to the two components with masses of $m_{\rm 1}$ and $m_{\rm 2}$, respectively. Both stars are assumed to rotate synchronously with the Keplerian angular frequency of $\Omega = \sqrt{G\left(m_1 +m_2\right)/a^3}$, where $a$ is the orbital separation.

The positions of the outer Lagrangian points ($x_{\rm L_{2}}$ and $x_{\rm L_{3}}$) for mass ratios $q < 1$ are obtained by fitting the Roche equipotential surface; we find
\begin{equation}
\label{eqs.6}
x_{\rm L_{2}}/a = -0.1617\,{(\log_{10}\,q)}^2 -0.2188\,\log_{10}\,q +1.1977 
\end{equation}
and
\begin{equation}
\label{eqs.7}
x_{\rm L_{3}}/a = -0.0559\,{(\log_{10}\,q)}^2 -0.2264\,\log_{10}\,q -1.1994.
\end{equation}

A small perturbation is sufficient to unbind particles located near the outer Lagrangian point \citep{2019MNRASHubova}. In our approach, the non-accreted material produced during the binary interaction is treated as ballistic particles launched from just outside the outer Lagrangian point closest to the compact companion \citep[e.g.][]{2007ARepSytov,2009ARepSytov,2019MNRASHubova}-- specifically, at a distance of $0.1a$ beyond the $L_2$ for NS companions, and $0.2a$ beyond the $L_3$ in the cases of BH companions (see Fig.\,\ref{fig: trajectory cartoons}). The launch position is set slightly farther from the binary system in the BH companion case, as the deeper gravitational potential well of the more massive component makes it harder for particles to escape. The particle is assumed to be initially at rest in the rotating frame at the launch position and the subsequent trajectory of each particle is obtained by numerically integrating its motion in the rotating reference frame. We follow the particle motion until it reaches a distance of approximately $100\,\rm AU$ from the binary system, at which point its velocity is regarded as its asymptotic velocity at infinity,
\begin{equation}
\label{eqs.8}
\vec{\upsilon}^{\rm RLOF} = \int_{0}^{\tau_{\rm R\sim 100\,\rm AU}} ( - \triangledown \psi \left(\,\vec{r} \,\right)- 2 \vec{\Omega} \times \vec{\upsilon})\,dt, 
\end{equation}
where the second term accounts for the Coriolis force encountered by a non-stationary particle in a co-rotating frame, and $\tau_{\rm R\sim 100\,\rm AU}$ is the time required for the particle to reach $100\,\rm AU$ from the system's centre of mass. The corresponding radial velocity in the inertial reference frame of an observer is then given by 
\begin{equation}
\label{eqs.9}
{\upsilon}_{\rm r}^{\rm RLOF} = \vec{\upsilon}^{\rm RLOF}  \cdot \hat{r}. 
\end{equation}

\begin{figure}
\begin{center}
\includegraphics[width=0.5\textwidth]{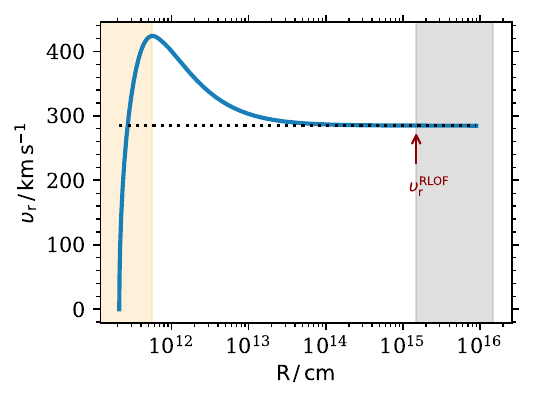}
\caption{Radial velocity evolution within $10\,\rm yr$ for the first particle ejected from the binary of $2.75M_{\odot}$ + NS. The dotted line indicates the horizon velocity determined at the end of the computation. The region of tidal acceleration is highlighted in orange, while the area beyond $100\,\rm AU$ is shaded in grey. The radial velocity at $100\rm AU$, denoted as $\upsilon_{\rm r}^{\rm RLOF}$, is considered as the particle’s asymptotic velocity. Beyond this distance, the particle moves at an approximately constant velocity of $\upsilon_{\rm r}^{\rm RLOF}$.}
\label{fig:radial velocity of one particle}
   \end{center}
\end{figure}

Here, this radial velocity at $R = 100\,\rm AU$, ${\upsilon}_{\rm r}^{\rm RLOF}$, is regarded as the asymptotic velocity of the outflow particle at infinity. This approximation is based on the dynamical behaviour of the outflow, and allows for a significant reduction in computational cost. As shown in Fig.\,\ref{fig:radial velocity of one particle}, the particle is rapidly accelerated within the first few orbital periods due to the tidal torque exerted by the binary system (highlighted in the orange colour), transferring angular momentum and energy from the binary orbit to the outflow \citep{1941ApJKuiper}. Subsequently, the particle experiences gravitational deceleration from the binary system, approximately proportional to $ - 1/R^2$. Beyond a distance of $100\rm\,AU$, the gravitational influence becomes negligible, and the radial velocity remains nearly constant, indicating that the particle eventually becomes unbound, and escapes to infinity with the asymptotic velocity of ${\upsilon}_{\rm r}^{\rm RLOF}$. 
In the example shown in Fig.\,\ref{fig:radial velocity of one particle}, the asymptotic velocity 
${\upsilon}_{\rm r}^{\rm RLOF} \sim 280\,\rm km\,s^{-1}$ corresponds to roughly $33\%$ of the escape velocity of the binary system. This outcomes is consistent with both ballistic calculations of outflows via the outer Lagrange point \citep{1979ApJShu} and the smoothed-particle radiation-hydrodynamics simulations \citep{2016MNRASPejcha}, which demonstrate that the asymptotic velocity of unbound Lagrangian outflows corresponds to a fraction of the binary escape velocity. 

We therefore assume that the particle continues to move outward with a constant velocity of ${\upsilon}_{\rm r}^{\rm RLOF}$ once it reaches $100\rm\,AU$ until the helium star collapses. Under this assumption, the radial distance reached by a particle at the time of core collapse is given by
\begin{equation}
\label{eqs.10}
R_{\rm RLOF} = 100{\rm AU} + {\upsilon}_{\rm r}^{\rm RLOF} \left ( \tau_{\rm rem} - \tau_{R\sim100\,\rm AU} \right).
\end{equation}
where $\tau_{\rm rem}$ denotes the remaining lifetime of the stripped helium star, and  $\tau_{R\sim100\,\rm AU} $ is the travel time required for the particle to reach $100\,\rm AU$.

Given the mass-transfer rate of $\dot{M}_{\rm RLOF}$ and the outflow velocity of $\upsilon_{\rm r}^{\rm RLOF}$, the density profile of the CSM formed by the RLOF outflow is expressed as:
\begin{equation}
\label{eqs.11}
\rho_{\rm RLOF} = \,\frac{1}{\eta}\frac{\dot{M}_{\rm RLOF}}{4 \pi {R^2_{\rm RLOF}} \upsilon_{\rm r}^{\rm RLOF}}.
\end{equation}
where $\eta$ represents the fraction of the full sphere occupied by the CSM. $\eta = 1$ corresponds to a spherically symmetric wind, while $\eta < 1$ represents a torus-like structure expected for the mass loss through outer Lagrange points \citep{2016MNRASPejcha, 2019MNRASHubova}. We consider both $\eta = 0.1$ and $\eta = 1$ when constructing the CSM density profiles in order to illustrate the effect of different CSM geometries on the circumstellar density distribution.

\subsection{Interaction-powered bolometric luminosity}
\label{sec: method interaction-powered luminosity}
The collision between the fast-moving SN ejecta and the slow-expanding dense CSM converts a fraction of the kinetic energy of the SN ejecta into radiation, thereby modifying the SN light curves. Assuming efficient radiative cooling, the rapidly expanding SN ejecta compress the dense CSM into a thin, dense shell in the shocked region. The bolometric luminosity powered by the forward shock, interacting with the swept-up CSM of density $\rho_{\rm csm}$ derived in Section \ref{sec:CSM density structures}, can be expressed as \citep{2013MNRASMoriya}:
\begin{equation}
\label{eqs.12}
L_{\rm inter} = 2 \pi \epsilon \rho_{\rm csm} r_{\rm sh}^2\upsilon_{\rm sh}^3,
\end{equation}
where $\epsilon$ denotes the efficiency of converting kinetic energy into radiation, for which we adopt a typical value of 0.1 \citep[e.g.][]{2013aMNRASMoriya}. For the luminosity calculation, a spherically symmetric density profile with $\eta = 1$ is applied, since the treatment of asymmetric structures with $\eta < 1$ introduces substantial complications, particularly regarding the absorption and viewing-angles effects. Here, $v_{\rm sh}$ and $r_{\rm sh}$ are the radius and velocity of the thin shock shell, respectively, both evolving as the shock propagates outward.

To determine the properties of the shock shell, we solve its dynamical evolution based on the conservation of mass and momentum as follows \citep{2013MNRASMoriya, 2025ApJWu}:
\begin{equation}
\label{eqs.13}
\frac{\it{d_{}} M_{\rm sh}}{\it{d} t} = 4 \pi r^2_{\rm sh} \left [ \rho_{\rm ej}\left( \upsilon_{\rm ej} - \upsilon_{\rm sh}\right) + \rho_{\rm csm} \left ( \upsilon_{\rm sh} - \upsilon_{\rm csm} \right )   \right ],
\end{equation}
\begin{equation}
\label{eqs.14}
M_{\rm sh} \frac{\it{d_{}} v_{\rm sh}}{\it{d} t} = 4 \pi r^2_{\rm sh} \left [ \rho_{\rm ej}\left( \upsilon_{\rm ej} - \upsilon_{\rm sh}\right)^2 + \rho_{\rm csm} \left ( \upsilon_{\rm sh} - \upsilon_{\rm csm} \right )^2   \right ].
\end{equation}
Here, at the shock radius of $r_{\rm sh}$, the velocity of the homogeneously expanding SN ejecta is  $v_{\rm ej} = r_{\rm sh}\,/\,t$. Following the numerical simulations of SN explosions by \citet{1999ApJMatzner}, a double power-law profile is applied to describe the structure of the SN ejecta, expressed as
\begin{equation}
\label{eqs.15}
\rho_{\rm ej}\left ( v_{\rm ej},t \right ) = \left
\{\begin{matrix}
\frac{1}{4\pi \left ( n - \delta  \right ) } \frac{A^{\left ( n-3 \right ) /2}}{B^{\left ( n-5 \right ) /2}} t^{-3} \upsilon^{-n}_{\rm ej}                       \hspace{0.5cm}      (\upsilon_{\rm ej} > \upsilon_t)   \\[0.1pt]
 \\
\frac{1}{4\pi \left ( n - \delta  \right ) } \frac{A^{\left ( \delta -3 \right ) /2}}{B^{\left ( \delta - 5 \right ) /2}} t^{-3} \upsilon^{-\delta}_{\rm ej}      \hspace{0.5cm}       (\upsilon_{\rm ej} < \upsilon_t) 
\end{matrix}\right.
\end{equation}
where 
$$ A =  2\left ( 5 - \delta \right ) \left ( n - 5 \right ) E_{\rm ej}$$
$$B =  \left ( 3 - \delta \right ) \left ( n - 3 \right ) M_{\rm ej}$$
$$\upsilon_t = (A/B)^{1/2}.$$
Ultra-stripped SNe are generally associated with strongly stripped progenitors in closed binaries \citep{2015MNRASTauris}, while observational ultra-stripped SN candidates are typically inferred to have low ejecta masses of a few tenths of a solar mass \citep{2024ApJDas}. The ejecta masses of most of our progenitor models fall within this regime (see Table\,\ref{tab:pre-after-MT}), and we therefore refer to them as ultra-stripped SN progenitors. We adopt a relatively low explosion energy of $E_{\rm ej} = 10^{50}\, \rm erg$ as the fiducial value \citep{2015MNRASTauris,2016MNRASLyman,2017MNRASMoriya,2018A&ATaddia}. For comparison, we also consider a higher explosion energy case with $E_{\rm ej} = 10^{51}\, \rm erg$. The SN ejecta mass from a progenitor with a final mass of $M_{1,f}$ is estimated as $M_{\rm ej} = M_{1,f} -  M_{\rm NS}$, assuming the remnant NS mass is $M_{\rm NS} = 1.4\,M_{\odot}$.

In addition to the interaction-powered luminosity, the contribution from the radioactive decay of $\rm ^{56}Ni$ is also included in our model. We adopt the diffusion model originally developed by \citet{1982ApJArnett} to analytically simulate how the deposited energy from radioactive decay is released from the homogeneously expanding SN ejecta (see also \citealt{2008MNRASValenti, 2009ApJChatzopoulos, 2012ApJChatzopoulos}). The luminosity at a specific time can be expressed as:
\begin{equation}
L_{\rm Ni}(t) = \frac{2}{t_{\mathrm{m}}}
\exp\!\left[-\left(\frac{t}{t_{\mathrm{m}}}\right)^{2}\right]
\int_{0}^{t}
\dot{Q}_{\mathrm{dep}}(t')
\exp\!\left[\left(\frac{t'}{t_{\mathrm{m}}}\right)^{2}\right]
\frac{t'}{t_{\mathrm{m}}}\,
\mathrm{d}t',
\label{eqs:16}
\end{equation}
in which the effective diffusion timescale is
\begin{equation}
\label{eqs.17}
t_m = \left(\frac{2\,\kappa_{\rm opt}M_{\rm ej}}{\beta c \upsilon_{\rm sc}}\right)^{1/2},
\end{equation}
where $\kappa_{\rm opt} = 0.06\,\rm cm^{2}\,g^{-1}$ is the effective optical opacity \citep{2003ApJMaeda,2016MNRASLyman,2011MNRASValenti}, c is the speed of light, and $\beta = 13.8$ is a structure constant \citep{2015MNRASWheeler,2016MNRASLyman}. Here, the characteristic expansion velocity of the SN is taken as $v_{\rm sc} = \sqrt{10E_{\rm ej}/3M_{\rm ej}}$, following \citet{2016MNRASLyman}. 

In Eq.~\eqref{eqs:16}, $\dot{Q}_{\mathrm{dep}}$ is the deposited energy from the radioactive decay that is thermalised within the ejecta and contributes to the bolometric luminosity. When the gamma-ray opacity decreases due to the ejecta expansion, gamma rays produced via radioactive decay escape from the SN ejecta without being thermalised, hence not contributing to the bolometric luminosity. This is most important for the late-time light curves of stripped-envelope SNe \citep{2015MNRASWheeler}. We take gamma-ray leakage into account following the formalism of \citet{1997ApJClocchiatti} and \citet{2015MNRASWheeler} with a characteristic timescale $T_0 = \left(C\kappa_{\gamma}M_{\rm ej}^2\,/\,E_{\rm ej}\right)^{1/2}$, where the opacity of the gamma rays is $k_\gamma = 0.03\,{\rm cm}^2\,\rm g^{-1}$ and C is a structure constant with a typical value of 0.05 \citep{2015MNRASWheeler}. The final deposited energy that can be thermalised is expressed as $\dot{Q}_{\rm dep} = \dot{Q}\left(1 - e^{-(T_0/t)^2} \right)$. The total radioactive energy generation rate $\dot{Q}$ is given by \citep{1994ApJSNadyozhin}
\begin{equation}
\label{eqs.18}
\dot{Q}(t) = 10^{43}\times\,(6.45\,e^{-t/8.8\,\rm day} + 1.45\,e^{-t/111.3\,\rm day})\,\frac{M(\rm ^{56}Ni)}{M_{\odot}}.
\end{equation}

The amount of $\rm ^{56}Ni$ synthesized in ultra-stripped supernovae remains uncertain. Observationally inferred nickel mass for ultra-stripped SN candidates are typically of order $\sim 0.02 - 0.05\,M_{\odot}$ \citep{2016MNRASLyman, 2018SciDe,2018A&ATaddia,2020ApJYao,2023ApJYan,2024ApJDas}, although values as low as a few $\sim 0.001\,M_{\odot}$ have also been reported for SN 2023zaw \citep{2024ApJDas}. Theoretical explosion calculations generally predict lower nickel yields and have difficulty reproducing the highest inferred values \citep{2015MNRASSuwa,2017MNRASYoshida,2018MNRASmuller,2022ApJSawada}. In this work, we adopt a representative value of $\rm M(^{56}Ni) \sim 0.01\,M_{\odot}$. The total bolometric luminosity is then $ L_{\rm bol}= L_{\rm inter} + L_{\rm Ni}$.

\subsection{Interaction-powered synchrotron emission}
\label{method: radio emission}

Across the entire electromagnetic spectrum, radio emission from SNe provides a crucial probe of the CSM surrounding their progenitors. Shock waves generated as the SN ejecta sweep up the dense CSM efficiently accelerate electrons, which then emit synchrotron radiation. The resulting luminosity at a frequency $\nu$ can be described approximately as \citep{1998ApJFransson,2004ApJBjornsson,2012ApJMaeda,sep2013A&AMoriya}
\begin{equation}
\label{eqs.19}
\nu L_{\nu} \approx \pi r^2_{\rm sh}\upsilon_{\rm sh}n_{\rm rel}
\left(\gamma_{\nu} \right)^{2-p} m_{\rm e}c^2
\left [ 1 +  \frac{t_{\rm sync}(\nu)}{t}\right ]^{-1}, 
\end{equation}
where $m_{\rm e}$ is the electron mass. The Lorentz factor of electrons radiating at the frequency $\nu$ in a magnetic field of strength $B$ is given by $\gamma_{\rm \nu} = (2\pi m_{\rm e} c \nu / eB )^{1/2}$, where $e$ is the electron charge. The number density of the relativistic electrons, $n_{\rm rel}$ , is assumed to follow a power-law energy distribution of $dn_{\rm rel}/ d\gamma \propto  \gamma^{-p}$, with $p = 3$ for Type Ib/c SNe \citep{2006ApJChevalier}. The synchrotron cooling timescale for electrons emitting at frequency $\nu$ is expressed as $t_{\rm sync} = 6 \pi m_{\rm e}c/(\sigma_T\gamma_\nu B^2)$, with $\sigma_T$ indicating the Thomson cross section.

Both synchrotron self-absorption and free-free absorption are taken into account in modelling the observable synchrotron emission. The synchrotron self-absorption optical depth is estimated as $\tau_{\rm ssa} = \left (\nu/\nu_{\rm ssa} \right)^{-(p+4)/2}$ with $\nu_{\rm ssa}\approx3\times10^5\left(r_{\rm sh} \epsilon_{\rm e}/\epsilon_{\rm B}\right)^{2/7}B^{9/7} \,\rm Hz$, implying that only high energy photons with frequencies of $\nu>\nu_{\rm ssa}$ can escape from the circumstellar environment. Here, $\epsilon_{\rm e}$ and $\epsilon_{\rm B}$ represent the fraction of the post-shock thermal energy that is used to accelerate the electrons and to amplify the magnetic field, respectively. $\epsilon_{\rm e} = 0.1$ and $\epsilon_{\rm B} = 0.01$ are applied to our models by default. The magnetic field strength here is parameterized as $B^2/8\pi = \epsilon_{\rm B}\,U_{\rm th}$, where $U_{\rm th}$ is the post-shock thermal energy density, calculated following equation~5 in \citet{2004ApJBjornsson}. Free-free absorption in the ionized CSM can further suppress the radio emission in the early stage after the SN explosion. The corresponding optical depth is $\tau_{\rm ffa} = \int_{r_{\rm sh}}^{\infty} \alpha_{\rm ffa}\,dr $ with an absorption coefficient \citep{1967ApJMezger,1979rpaRybicki} of $\alpha_{\rm ffa} = 3.8  \times 10^{-29} \rm cm^{-1} \it \left (n_e \sum{n_i Z^{\rm 2}} \right)_{\rm csm} \times \left (\frac{T_{\rm e,csm}}{\rm 10^5\, K}\right )^{\rm -1.35} \left (\frac{v}{\rm 10\, GHz} \right )^{\rm -2.1}$. Here, we assume an electron temperature in the CSM of $T_{\rm e,csm}\sim 10^5\,\rm K$. For a fully ionized, helium-rich gas, the electron number density is given by $n_e = \rho_{\rm csm}/2m_p$ and the ion number density with charge of $Z = 2$ is $n_i =\rho_{\rm csm}/4m_p$, respectively, where $m_p$ denotes the proton mass. The observed spectrum is then obtained by accounting for these absorption processes:
\begin{equation}
\label{eqs.20}
L_{\rm \nu,obs} = L_{\rm \nu} e^{-\tau_{\rm ffa}} \left(\frac{1-e^{-\tau_{\rm ssa}}}{\tau_{\rm ssa}}\right).
\end{equation}

\begin{table*}[h]
    \renewcommand  
    \arraystretch{1.5} 
    \centering
    \setlength\tabcolsep{10.0pt}
    \caption{Initial and pre-SN properties of the binary models and the resulting SN ejecta and CSM properties.}
    \begin{threeparttable}  
    \begin{tabular}{c c c c c c c c c c c c c c c}
    \hline 
    \hline 
     Binary systems            & $M_{\rm 1,i}$     & $P_{\rm orb,i}$ &  $M_{\rm core,i}^{\rm CO}$ & $M_{\rm 1,f}$     & $P_{\rm orb,f}$  &  $M_{\rm core,f}^{\rm CO}$  & $M_{\rm He,f}^{\rm env}$  & $\Delta M_{\rm acc}$ & $M_{\rm ej}$  & $M_{\rm CSM}$  \\ 
     & $(\rm M_{\odot})$      & $(\rm days)$    &  $(\rm M_{\odot})$ &   $(\rm M_{\odot})$  & $(\rm days)$     &  $(\rm M_{\odot})$  &  $(\rm M_{\odot})$ & $(\rm M_{\odot})$ &  $(\rm M_{\odot})$ & $(\rm M_{\odot})$    \\ 
    \hline  
    2.85\,$\rm M_{\odot}$ + BH & 2.85 & 4.39 & 1.45 & 2.47 & 5.84 & 1.57 & 0.90 & 1.05e-2 & 1.07 & 0.38   \\
    
    2.85\,$\rm M_{\odot}$ + NS & 2.85 & 1.25 & 1.45 & 1.86 & 0.94 & 1.57 & 0.29 & 1.00e-3 &     0.46 & 0.99  \\    
    2.80\,$\rm M_{\odot}$ + BH & 2.80 & 1.98 & 1.45 & 2.05 & 3.88 & 1.56 & 0.49 & 9.66e-3 &     0.65 & 0.75  \\ 
    
    2.80\,$\rm M_{\odot}$ + NS & 2.80 & 0.53 & 1.45 & 1.79 & 0.42 & 1.56 & 0.23 & 1.00e-3 &  0.39 & 1.01  \\
    
    2.75\,$\rm M_{\odot}$ + BH & 2.75 & 0.74 & 1.45 & 1.88 & 1.69 & 1.56 & 0.32 & 2.69e-3 & 0.48 & 0.87  \\ 
    
    2.75\,$\rm M_{\odot}$ + NS & 2.75 & 0.19 & 1.45 & 1.66 & 0.16 & 1.54 & 0.12 & 4.96e-4 & 0.26 & 1.09  \\     
    \hline 
    \end{tabular}
        \begin{minipage}{0.90\textwidth}
         \footnotesize                       
         \textit{Notes.}
            Columns (2)--(4) list the initial mass ($M_{\rm1,i}$), orbital period ($P_{\rm orb,i}$), and CO-core mass ($M_{\rm core,i}^{\rm CO}$) of the stripped-star progenitor. Columns (5)--(8) give the corresponding pre-SN properties, including the final stripped-star mass ($M_{\rm1,f}$), orbital period ($P_{\rm orb,f}$), CO-core mass ($M_{\rm core,f}^{\rm CO}$), and He-rich envelope mass ($M_{\rm He,f}^{\rm env}$). $\Delta M_{\rm acc}$ denotes the total mass accreted by the compact companion. $M_{\rm ej}$ is the SN ejecta mass assuming a $1.4\,M_\odot$ NS remnant, and $M_{\rm CSM}$ is the total mass lost from the binary system, representing an upper limit to the CSM mass. The NS and BH companion masses are fixed at $1.4\,M_\odot$ and $5.0\,M_\odot$, respectively.
        \end{minipage}           
    \end{threeparttable}       
    \label{tab:pre-after-MT}
\end{table*}

\begin{figure*}
\begin{center}
\includegraphics[width=1.0\textwidth]{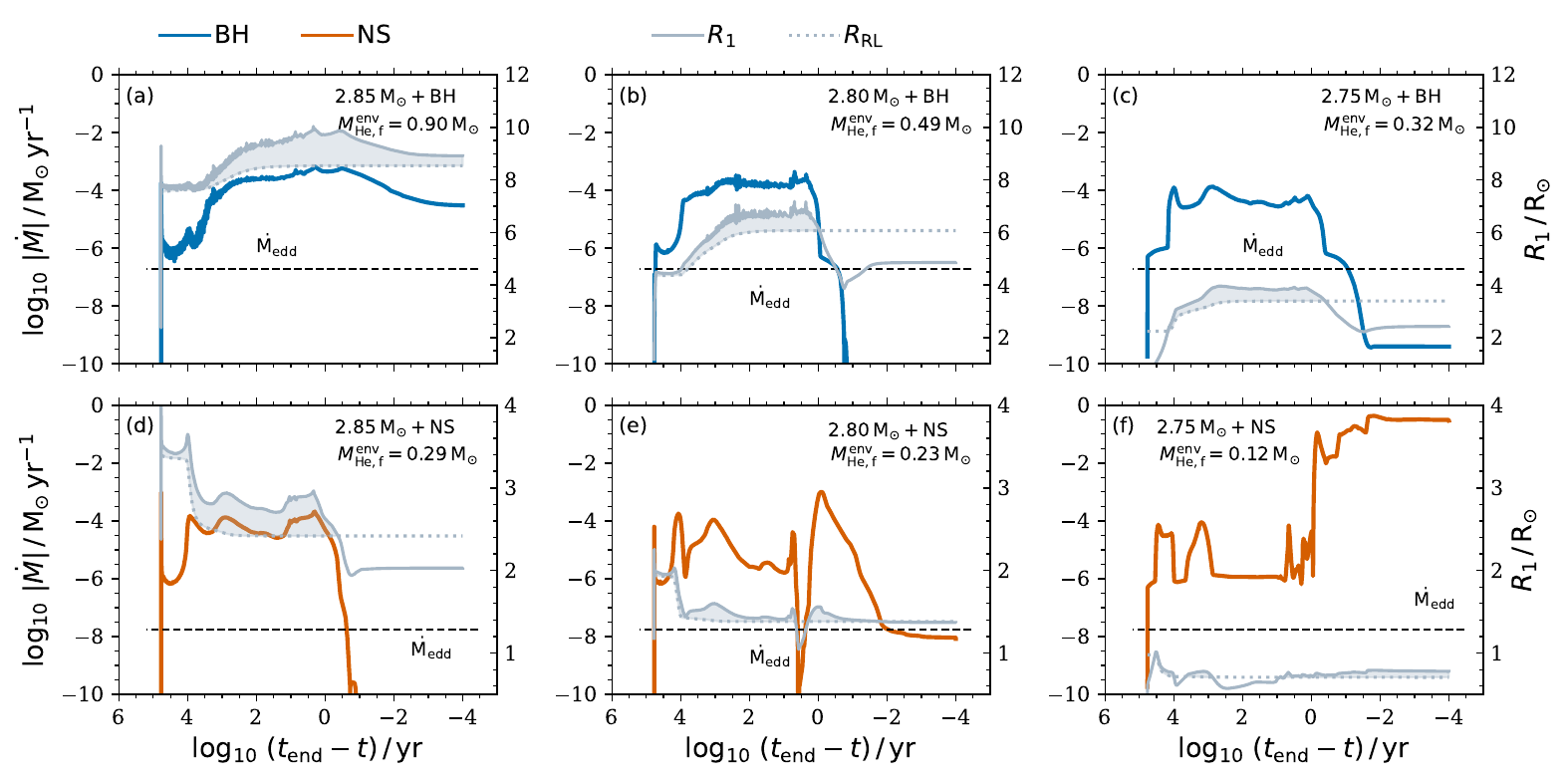}
\caption{Mass-loss rates and (Roche-lobe) radii of the stripped helium stars orbiting around compact (BH or NS) companions with the time evolution until Si depletion inside the core. The mass-loss rates of systems with a BH or NS companion are shown by the blue (panel a, b and c) and orange lines (panel d, e and f), respectively. In each panel, the grey solid and dotted lines indicate the stellar surface radii and Roche-lobe radii of the stripped helium star, respectively. The grey-shaded regions mark the phases during which the stellar surface exceeds its Roche-lobe radius. The Eddington-limited accretion rates for the NS and BH companion are $\sim \rm 1.8 \times 10^{-8} {M}_{\rm \odot}$ and $\sim \rm 1.9 \times 10^{-7} {M}_{\rm \odot}$, respectively, which is shown by the dashed black lines.}
\label{fig:mass_dot and radii wind}
   \end{center}
\end{figure*}

\section{Results}
\label{results part}

Mass-loss histories of different binary systems are presented in Sect.\,\ref{sec:Binary models}, with detailed model parameters summarised in Table\,\ref{tab:pre-after-MT}. In Sect.\,\ref{sec:Mirror effect}, the mirror principle is applied to interpret how the internal energy generation drives the variations in the surface radius during the donor star's evolution. Given the mass-loss rates from the binary systems, the resulting CSM density structures surrounding the distinct SN progenitors are investigated in Sect.\,\ref{sec:CSM density structures}. The interaction-powered light curves, including the bolometric and radio light curves, are discussed in Sect.\,\ref{sec:Interaction-powered luminosity}.

\subsection{Mass loss from the binary models before core collapse}
\label{sec:Binary models}

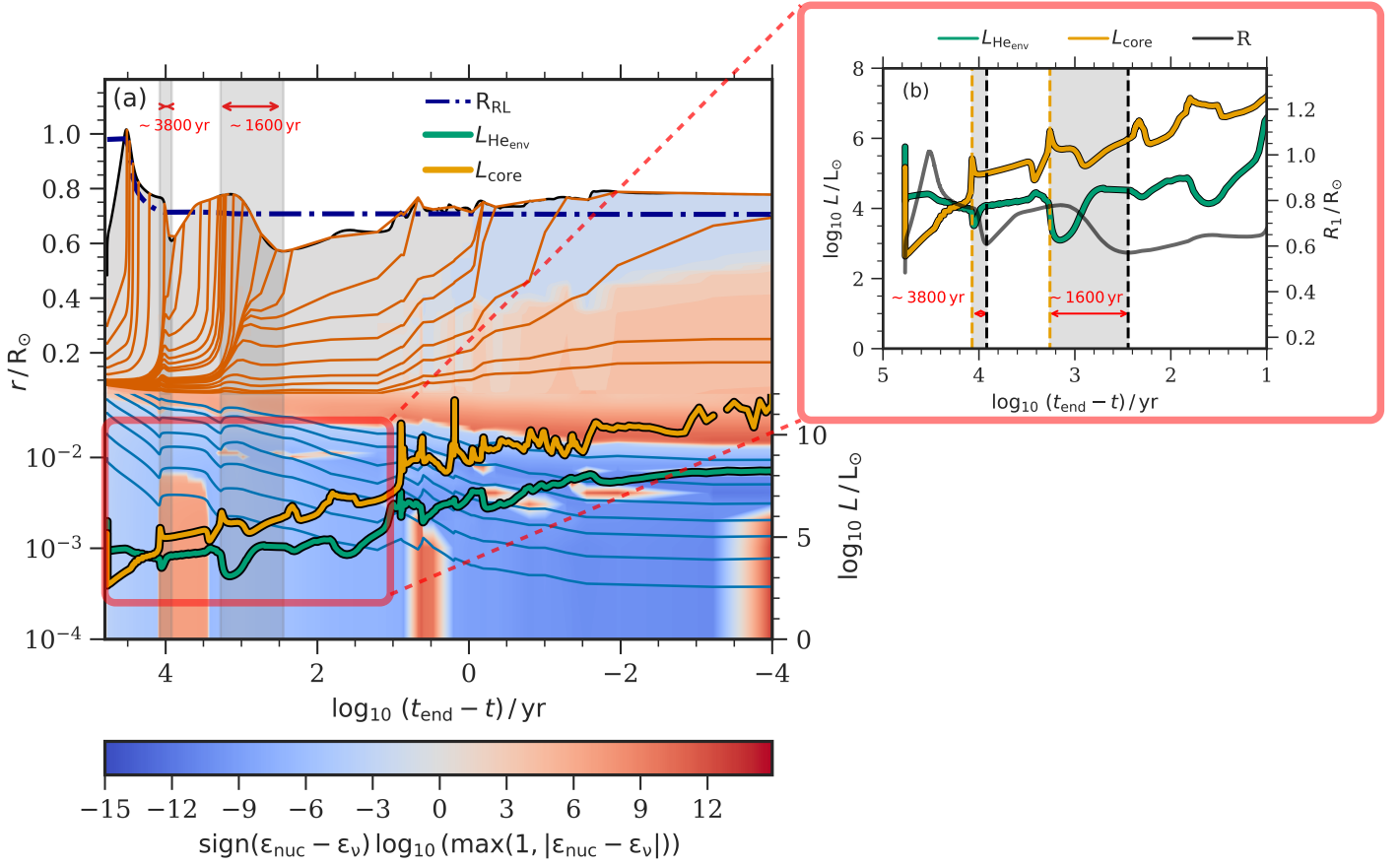
\begin{figure*}[t]
    \begin{tikzpicture}

        \node[anchor=south west,inner sep=0] (main) at (0,0) {
            \includegraphics[width=0.65\textwidth]{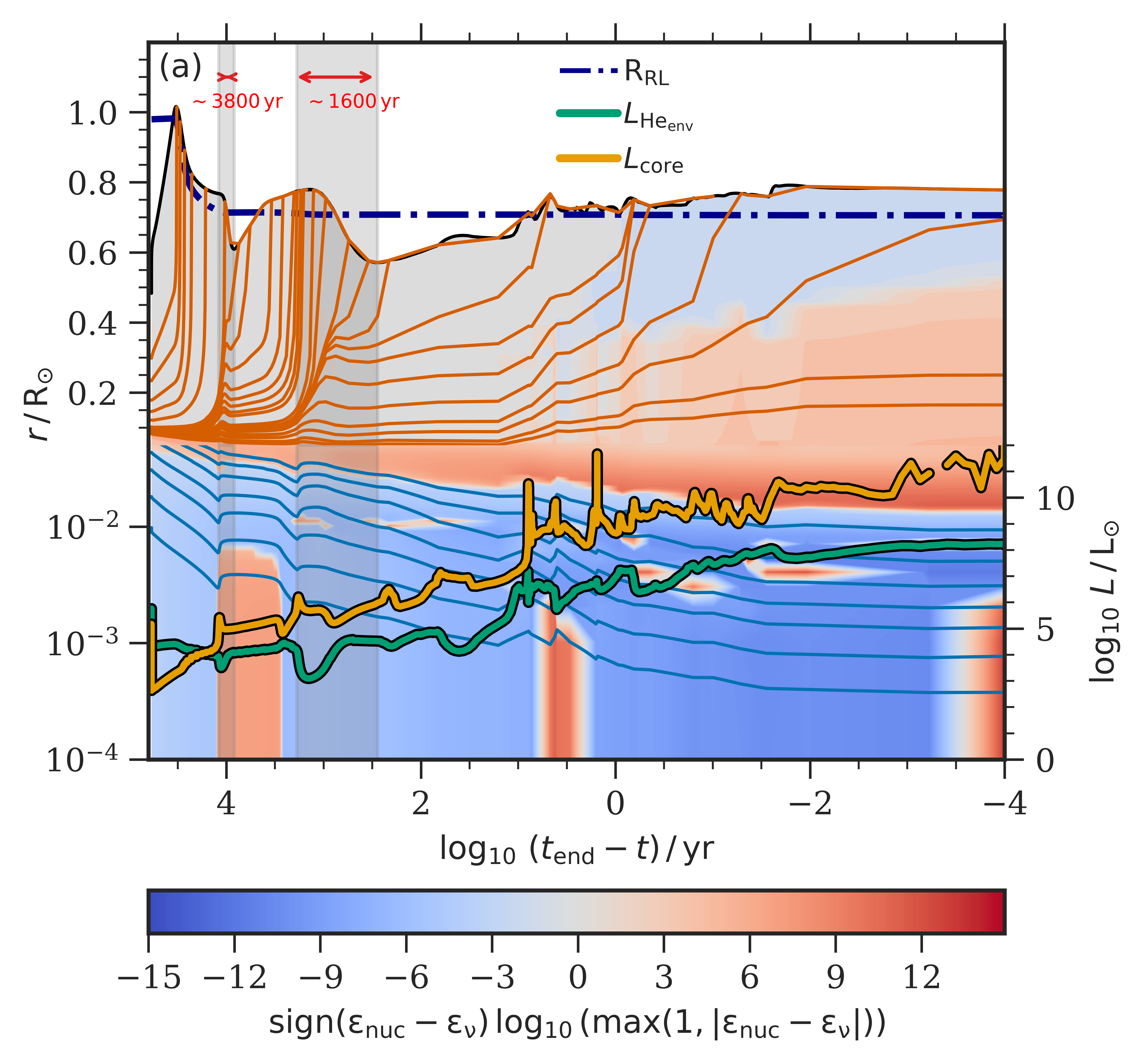}
        };

        \begin{scope}[x={(main.south east)}, y={(main.north west)}]  

            \draw[red, thick, rounded corners, line width = 1.0mm, opacity=0.5] (0.13, 0.33) rectangle (0.45, 0.55);

            \draw[red, thick, opacity=0.7, line width = 0.5mm, dashed] (0.45, 0.55) -- (0.91, 1.05);
            \draw[red, thick, opacity=0.7, line width = 0.5mm, dashed] (0.45, 0.34) -- (0.91, 0.55);

            \draw[red, thick, rounded corners, line width = 1.0mm, opacity=0.5] (0.91, 0.55) rectangle (1.56, 1.05);
        \end{scope}

        \node[anchor=south west, inner sep=0] (zoom) at (11, 6) {
            \includegraphics[width=0.4\textwidth]{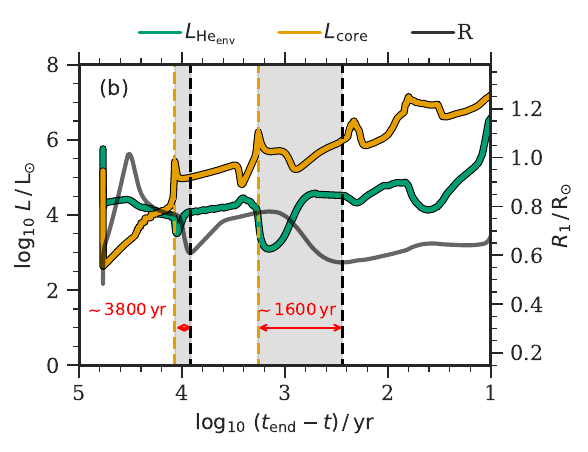}
        };

    \end{tikzpicture}
    \caption{Time evolution of the stellar structure of $2.75\,M_{\odot}$ stripped helium star in radius coordinate, orbiting a NS companion. Colors here denote the net energy between the nuclear burning and the neutrino cooling, where $\epsilon_{\rm nuc}$ and $\epsilon_{\nu}$ denote the specific nuclear energy generation and neutrino loss rates, respectively. Zones dominated by nuclear burning are indicated by red colors with positive values, while blue colors with negative values represent regions which are dominated by neutrino cooling. Blue and orange solid lines indicate the mass shell in the core and envelope, respectively. In both panels (a) and (b), the accumulated luminosities within the core and envelope are shown by the yellow and green solid lines, respectively. Grey shaded regions indicate the time delay required for the surface radius to respond to the inner nuclear generation. The dashed yellow and black lines mark the peak core luminosity and minimum surface radius during a given nuclear-burning phase, while the black solid line shows the evolution of the surface radius.}
    \label{fig:Kipp_mirror_effect}
\end{figure*}

A considerable amount of material is ejected from the progenitor system during the binary interactions before core collapse, potentially contributing to the surrounding CSM that interacts with the subsequent SN ejecta. Fig.\,\ref{fig:mass_dot and radii wind} shows the mass-loss rates of the stripped helium stars in different binary configurations, including both the wind mass loss and the RLOF outflow. As shown in Fig.\,\ref{fig:mass_dot and radii wind}, the majority of the transferred material can not be accreted by the compact companion, as the mass-transfer rates exceed the corresponding Eddington limits (indicated by the black dotted lines) by about $3-4$ orders of magnitude. Binary systems with shorter orbital periods contribute more mass to the CSM, as shown in Table\,\ref{tab:pre-after-MT}. For instance, around $1.09\,\rm M_{\odot}$ He-rich material is ejected from the binary system with a $2.75\,\rm M_{\odot} \rm$ helium star + $\rm NS$ and initial $0.19\,\rm d$ orbit, while only $\sim 0.38\,\rm M_{\odot}$ He-rich material contributes to the CSM in the case of a $2.85\,\rm M_{\odot}$ helium star + $\rm BH$ with a larger initial orbital period of $\sim 4.39\,\rm d$. In total, the unaccreted material, with a total mass of up to $\sim 1.0\,\rm M_{\odot}$ (see $M_{\rm CSM}$ in Table\,\ref{tab:pre-after-MT}), escapes the binary system and is available to form the CSM surrounding the progenitor before the SN explosion.

Variations in the mass-transfer rate are driven by the changes in the surface radii of the stripped helium stars during their late evolutionary stages. The differences between the surface radius and the Roche-lobe radius set the RLOF mass loss during the binary interaction. As shown in Fig.\,\ref{fig:mass_dot and radii wind}, the Roche lobe radii (indicated by grey dotted lines) remain nearly constant in the final few thousand years before core collapse. This is because the binary configuration -- such as the mass ratio and separation -- does not change significantly over such a short remaining timescale. Under these conditions, the evolution of the stellar radius directly determines the corresponding mass-transfer (or mass-loss) rate of the binary system. This behaviour is consistently visible across different binary models in Fig.\,\ref{fig:mass_dot and radii wind}.

In addition, we find that stripped helium stars with less massive final He-rich envelopes tend to be more compact and reveal more variable mass-loss histories. For instance, the stripped helium star of $2.75\,\rm M_{\odot}$ orbiting a NS companion retains a final He-rich envelope of only $\sim 0.12\, \rm M_{\odot}$, which is much smaller than that for the $2.85\,\rm M_{\odot}$ helium star with a BH companion ($M_{\rm He,f}^{\rm env} \sim 0.90\,\rm M_{\odot}$). 
Correspondingly, the former binary system, with its more compact envelope, shows multi-peaked mass-loss behaviour (see Fig.\,\ref{fig:mass_dot and radii wind}{\color{blue}f}), in contrast to the continuous and smoother mass-loss profile of the latter system (shown in the Fig.\,\ref{fig:mass_dot and radii wind}{\color{blue}a}). Investigating the physical mechanisms that drive the surface-radius variations from the stellar-evolution perspective is essential for understanding the RLOF mass-loss processes that ultimately shape the CSM around SN progenitors.

\subsection{Mirror principle: surface-radius variations in response to the internal energy generation}
\label{sec:Mirror effect}

\begin{figure*}[htbp]
\centering
\begin{minipage}[c]{0.69\textwidth}
  \includegraphics[width=\textwidth]{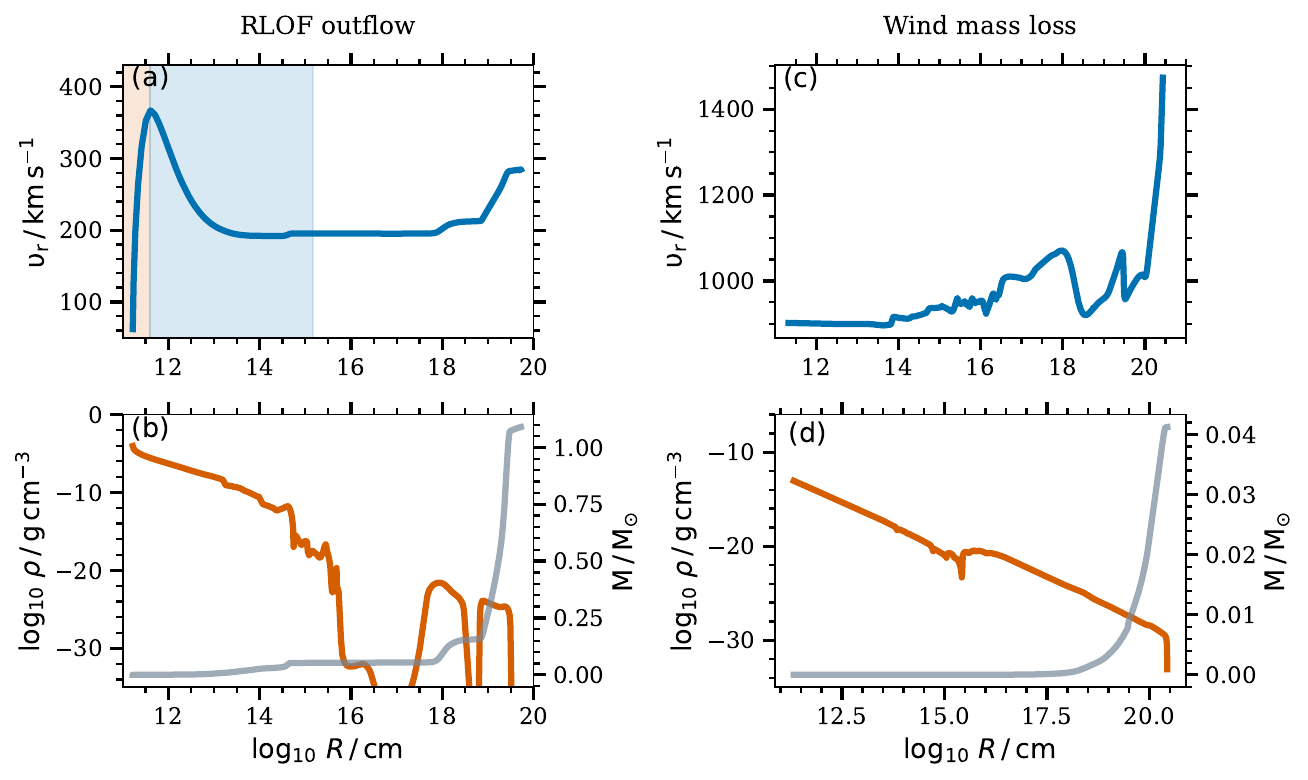}
\end{minipage}
\hfill
\begin{minipage}[c]{0.3\textwidth}
  \caption{Radial velocity, density structure, and accumulated mass of the material lost from the binary system via the RLOF outflow (panels a and b) or the stellar wind (panels c and d), plotted as a function of the travel distance at the time of core collapse. The solid blue and orange lines indicate the radial velocity and the density structure, respectively. The solid grey line denotes the cumulative mass lost through the corresponding mass-loss channel. Here, the RLOF outflow is confined within a torus-like CSM with a geometric factor of $\eta = 0.1$, while the mass lost via stellar winds is assumed to form a spherically symmetric CSM structure. In panel a, the orange-shaded region marks material experiencing tidal acceleration, while the blue-shaded region corresponds to material undergoing gravitational deceleration.  
  }
  \label{fig:density_profile_individial}
\end{minipage}
\end{figure*}

To explore the physical origin of the surface radius variations, we take the stripped helium star of $2.75\,M_{\odot}$ orbiting a NS companion as an example and look into its interior; this helium star exhibits the most complex and diverse radius variation, and consequently, the strongest variations in the mass-transfer rate. In this section, we illustrate how the surface radius of a stripped helium star with a thin helium envelope responds to internal nuclear fusion with a corresponding thermal-timescale delay.

In Fig.\,\ref{fig:Kipp_mirror_effect}{\color{blue}a}, we show how the nuclear energy generated in the core propagates outward by tracing the evolution of the various mass shells. The blue and orange solid curves indicate the mass contours within the core and the envelope, respectively. The increasing and decreasing trends of a given mass shell in radius coordinate correspond to its expansion and contraction, respectively. The initial expansion of the surface radius, which triggers the first mass-transfer episode, results from the thermal adjustment of the stripped helium star in response to the previous rapid envelope removal in the CE phase (see \citealt{Wei2024A&A}). Similar behaviour has also been reported in other stripped helium-star models \citep{2022MNRASVigna}. In addition to the mass contours, the nuclear energy generation in the core and helium envelope is presented by yellow and green solid lines, respectively. The carbon-oxygen (CO) core contracts until it becomes sufficiently hot to reach core C ignition. During this contraction, the envelope expands because of the mirror principle \citep[the helium-burning shell acts as the mirror;][]{1990ssebookKippenhahn}. Once core C burning starts, the core expands, and the core luminosity reaches a peak value. This reduces the helium-burning shell luminosity, leading to a gradual shrinking of the stellar surface radius. Once the shrinking surface radius becomes smaller than the Roche lobe radius, the star detaches from its Roche lobe, thereby terminating the first mass-transfer episode. Three distinct mass-transfer episodes can be identified in Fig.\,\ref{fig:Kipp_mirror_effect}{\color{blue}a} by comparing the stellar surface radius with its Roche-lobe radius. In the following, we connect the surface-radius evolution to changes in nuclear burning inside the core.

As shown in Fig.\,\ref{fig:Kipp_mirror_effect}{\color{blue}b}, the minimum stellar surface radius (left dashed black line) occurs after the core reaches its peak luminosity (left dashed yellow line). By following the time evolution of energy generation in the core and the stellar surface radius, we find that the stellar surface does not respond instantaneously to the nuclear burning within the core. Instead, there is a delay of approximately $3800\,\rm yr$ before the surface of the helium star responds to the enhanced core energy release, as indicated by the first grey-shaded region in the Figs.\,\ref{fig:Kipp_mirror_effect}{\color{blue}a} and {\color{blue}4b}. Notably, the time delay is consistent with the corresponding thermal (Kelvin-Helmholtz) timescale, which can be estimated as $\tau_{\rm KH} \approx \left(GM_{1}\Delta M\right)/\left(2R_{1}L_{1}\right) \sim 3000\,\rm yr$, where $M_1$ and $\Delta M$ denote the total stellar mass and the mass above the nuclear-burning zone, respectively, and $R_1$ and $L_1$ are the stellar surface radius and luminosity. 
 
In addition to central C burning, C-shell burning also induces variations in the stellar surface radius, but with a shorter time delay. The second grey-shaded region in Figs.\,\ref{fig:Kipp_mirror_effect}{\color{blue}a} and {\color{blue}4b} corresponds to the first C-shell burning episode and reveals a slightly shorter delay time of approximately $1600\,\rm yr$, owing to the smaller amount of overlying mass above the shell-burning zone compared to the center-C burning region. In principle, the stellar surface radius is expected to respond to each internal burning episode -- whether central or shell burning -- with a thermal-timescale delay, provided that the delay time is shorter than the remaining lifetime of the star. However, after central C depletion, it becomes difficult to follow this behaviour in detail. In the final $\sim10\,\rm yr$ before collapse, the stellar envelope expands substantially due to enhanced He-shell burning, leading to extremely high mass-transfer rates. 

The surface-radius variations during the late-stage evolution of the star retain imprints of the nuclear reaction occurring in the core, though with a thermal-timescale delay. Notably, not all stripped-helium stars are capable of reflecting the internal nuclear activity through observable surface-radius variations. As shown in Fig.\,\ref{fig:mass_dot and radii wind}, the surface of the stripped helium star with less massive envelopes are more sensitive to ``feel'' what happened in their cores, exhibiting more complex and diverse structures in the corresponding mass-transfer rates. For more massive envelopes, the radius variations are more easily absorbed inside the envelopes and do not show up at the surface. These variations, in turn, influence the density structure of the CSM surrounding the SN progenitors (Sect.\,\ref{sec:CSM density structures}) and consequently affect the properties of the interaction-powered light curves (Sect.\,\ref{sec:Interaction-powered luminosity}).

\begin{figure*}[t]
    \centering
    \begin{tikzpicture}[x={(1cm,0cm)}, y={(0cm,1cm)}]
       
        \node[anchor=west,inner sep=0, xshift=1cm] (main) at (0.0,0) {
            \includegraphics[width=0.45\textwidth]{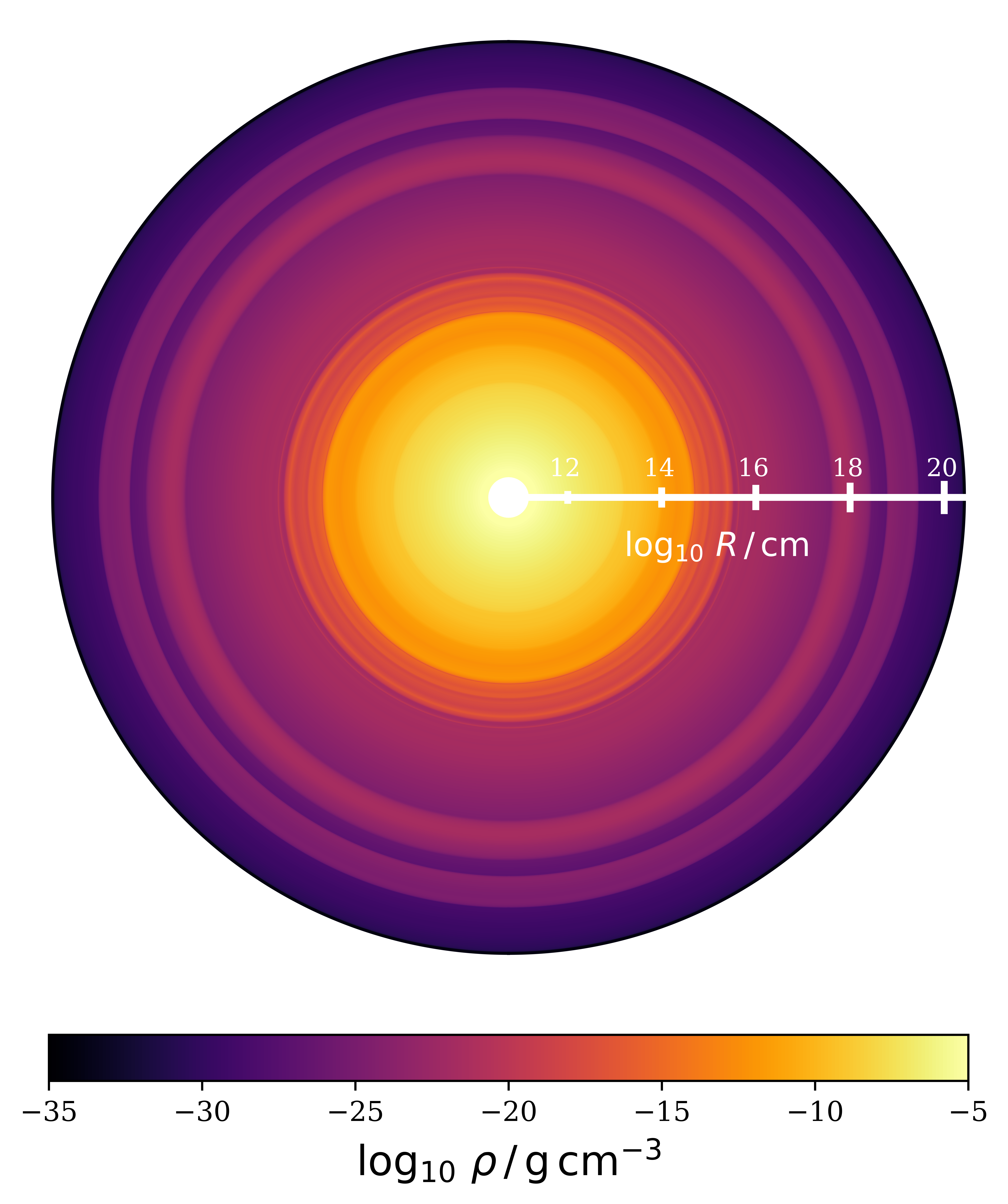}
        };

        \begin{scope}[x={(main.south east)}, y={(main.north west)}]  
           
            \draw[black, thick, opacity=0.7, line width = 0.5mm, dashed] (0.5, 0.68) -- (1.06, 0.35);
            \draw[black, thick, opacity=0.7, line width = 0.5mm, dashed] (0.85, 1.03) -- (1.06, 1.05);

        \end{scope}

        \node[anchor= west, inner sep=0] (zoom) at (10, 0) {
            \includegraphics[width=0.38\textwidth]{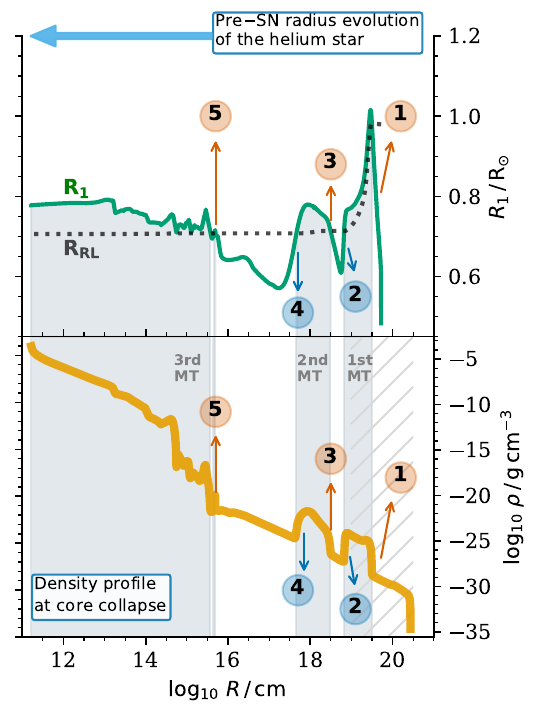}
        };

    \end{tikzpicture}
    \caption{Overall CSM density structure surrounding a $2.75\,M_{\odot}$ helium star in orbit with a NS companion before its core collapse, constructed from the density contribution shown in Fig.\,\ref{fig:density_profile_individial}. The left panel shows the density distribution in the equatorial plane of the torus-like CSM ($\eta = 0.1$), where the colour indicates the density. The bottom-right panel presents the corresponding radial density profile at core collapse (yellow line). The top-right panel displays the evolution of the stellar surface radius ($R_{\rm 1}$) and its Roche-lobe radius ($R_{\rm RL}$), shown by the green solid and grey dotted lines, respectively. The three mass-transfer epsiodes (1st – 3rd MT), during which $R_{\rm 1} > R_{\rm RL}$, are highlighted in light grey. The grey hatched region indicates the radial range ($\rm log\,R\,/\,cm>19$) where interaction with interstellar medium may modify the density structure.}
    \label{fig:overall_density_2_75_NS}
\end{figure*}

\subsection{CSM density structures before collapse}
\label{sec:CSM density structures}

\begin{figure}
\begin{center}
\includegraphics[width=0.5\textwidth]{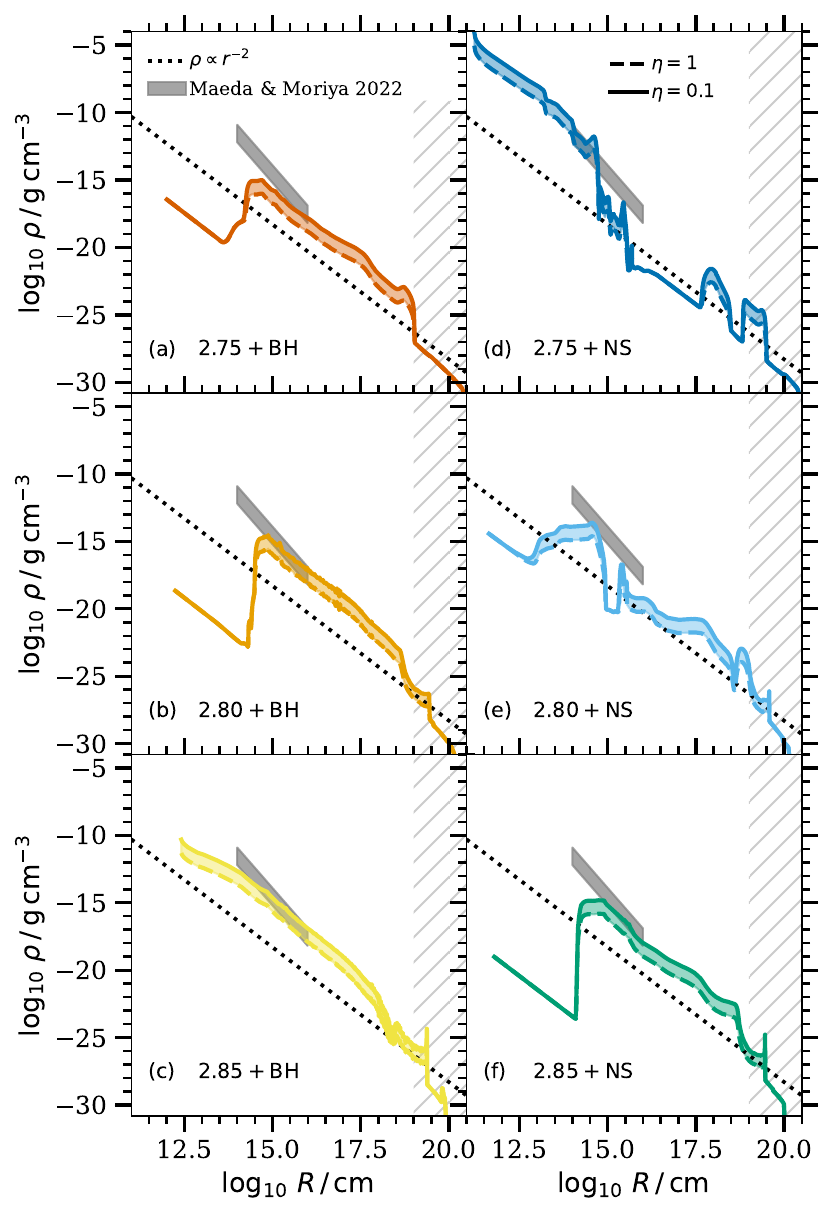}
\caption{CSM density profiles derived from the mass-loss histories of the binary systems listed in Table\,\ref{tab:pre-after-MT}. For each model, the solid line indicates material confined within a torus-like CSM with a geometric covering factor of $\eta = 0.1$, while the dashed lines correspond to isotropic CSM distributions. The black dotted lines show the CSM density expected from a typical Wolf-Rayet wind, assuming a mass-loss rate of $10^{-5} \rm M_{\odot} yr^{-1}$ and a velocity of $10^3\,\rm km\,s^{-1}$. The grey-shaded region marks the range of CSM densities inferred from the SNe Ibn observations in \citet{2022ApJMaeda}. The grey hatched region indicates the radial range ($\log\,R\,/\,\rm cm >19$) where interaction with interstellar medium may modify the density structure.}
\label{fig:density profile from all binaries}
   \end{center}
\end{figure}

Material escapes from the binary system through two primary channels: stellar winds and RLOF outflow. To investigate the differences in the properties of the material ejected through these channels (e.g. mass-loss rate and velocity), we adopt a fiducial model consisting of a $2.75\,\rm M_{\odot}$ helium star orbiting a NS companion. We further explore how the material from each channel contributes to the CSM structure at the time of core collapse.

The RLOF outflow is the dominant channel through which material escapes the binary system. During binary interaction, the mass-transfer rate exceeds the wind mass-loss rate by at least two orders of magnitude. Consequently, nearly $1.00\,\rm M_{\odot}$ of He-rich material is lost through the RLOF outflow (see Fig.\,\ref{fig:density_profile_individial}{\color{blue}b}), while only $0.04\,\rm M_{\odot}$ is lost via wind mass loss (see Fig.\,\ref{fig:density_profile_individial}{\color{blue}d}). Material ejected via these two channels exhibits different velocities and mass-loss histories, which together shape the final CSM density profile at the time of explosion. Fig.\,\ref{fig:density_profile_individial}{\color{blue}a} shows the radial velocity of the RLOF outflow at core collapse. Material within $\sim 10^{11.5}\rm cm$ remains subject to the binary tidal acceleration, while material at larger distances experiences gravitational deceleration. Beyond a distance of roughly $100\,\rm AU \sim 10^{15}\rm cm$, the outflow reaches its asymptotic velocity. Variations in these asymptotic velocities arise from the changes in the binary escape velocities during the binary evolution (see the discussion in Sect.\,\ref{sec: method RLOF mass loss}). In total, the RLOF outflow expels about $1.0\,\rm M_{\odot}$ of material (solid grey line in Fig.\,\ref{fig:density_profile_individial}{\color{blue}b}), producing a complex, non-monotonic density structure at the time of core collapse (solid orange line in Fig.\,\ref{fig:density_profile_individial}{\color{blue}b}). Here, the RLOF outflow is assumed to be confined within a torus-like structure with a geometric covering factor of $\eta = 0.1$. The variability of the mass-transfer rate is a key factor in shaping the non-monotonic density structure of RLOF outflow. Stellar winds provide the second channel for mass loss, with velocities comparable to the surface escape speed of the stripped helium star. As shown in Fig.\,\ref{fig:density_profile_individial}{\color{blue}c}, the wind velocity can reach $\sim1000\,\rm km\,s^{-1}$, nearly an order of magnitude higher than that of the RLOF outflow (Fig.\,\ref{fig:density_profile_individial}{\color{blue}a}). However, because only $0.04\,\rm M_{\odot}$ of material is lost through winds (Fig.\,\ref{fig:density_profile_individial}{\color{blue}d}), the resulting wind-formed CSM is significantly less dense and display a simpler structure (Fig.\,\ref{fig:density_profile_individial}{\color{blue}d}).

The overall CSM density structure at the moment of core collapse results from the combined contributions of both mass-loss channels, as shown in Fig.\,\ref{fig:overall_density_2_75_NS}. The CSM density reaches up to $\sim 10^{-5}\,\rm g\,cm^{-3}$ at the inner edge, and decreases outward by more than twenty orders of magnitude. Several prominent peaks appear at radii of $\sim 10^{15}\,\rm cm$, $10^{18}\, \rm cm$ and $10^{19}\, \rm cm$ (bottom-right panel in Fig.\,\ref{fig:overall_density_2_75_NS}), reflecting variations in the total mass-loss rate driven by the late-stage radius evolution of the strippped helium star. Comparing the radius evolution (top-right panel) with the CSM density profile (bottom-right panel) reveals a clear correspondence between surface expansion or contraction and the resulting CSM density variations. The initial expansion (labelled “1”), caused by the thermal relaxation in response to the rapid envelope removal, triggers the first rise in the density profile (also labelled “1”). According to the mirror principle, decreasing energy generation in the helium-burning shell follows the central carbon ignition (Fig.\,\ref{fig:Kipp_mirror_effect}), leading to the surface contraction (labelled “2”) and its detachment from its Roche-lobe radius. As a result, the first density peak forms due to the first mass-transfer episode. The subsequent recurrence of the energy generation in the helium-burning shell drives another surface expansion (labelled “3”), which triggers the second mass-transfer episode. The ensuing carbon-shell burning reduces the helium-burning shell luminosity, causing the surface radius to contract again (labelled “4”). Consequently, the second mass-transfer phase ceases, leaving behind the corresponding second density peak. Finally, a pronounced surface expansion shortly before core collapse triggers the third mass-transfer episode, generating the last rise in the density profile (labelled “5”). Therefore, the resulting CSM structure preserves valuable information about the late-stage evolution of the stripped helium star (particularly its nuclear activity and surface radius variations), as well as the role of binary interaction in shaping its mass-loss history. However, the material located beyond $\rm \,R > 10^{19}\,\rm cm$ may interact with the surrounding interstellar medium \citep{1977ApJWeaver, 2022ApJMatsuoka}, as indicated by the shaded region in Fig.\,\ref{fig:overall_density_2_75_NS}. Such interactions can modify the outer CSM structure and potentially lead to a density enhancement. Therefore, the density in this region may be higher than that predicted by our binary-evolution model.

Mass-loss processes in binaries containing distinct stripped helium stars lead to different CSM density structures at the time of core collapse. As shown in Fig.\,\ref{fig:density profile from all binaries} (with corresponding two-dimensional CSM density maps presented in Fig.\,\ref{fig:density_structure}), the overall density decline generally follows the black dotted line of $\rho \sim r^{-2}$. The model shown in Fig.\,\ref{fig:density profile from all binaries}{\color{blue}c} exhibits a continuous wind-like CSM structure with densities 3--4 orders of magnitude higher than those expected from typical Wolf-Rayet winds. Some models produce detached CSM configurations (e.g. models in Figs.\,\ref{fig:density profile from all binaries}{\color{blue}a, 7b} and {\color{blue} 7f}), whereas others display more complex and non-monotonic features, particularly systems with lower-mass He-rich envelopes (e.g. Figs.\,\ref{fig:density profile from all binaries}{\color{blue}d} and {\color{blue} 7e}). This diversity arises because the stripped helium stars with thinner envelopes experience more complex binary mass-transfer episodes, as described in Sect.\,\ref{sec:Mirror effect}. The resulting CSM densities predicted by our models can approach the density regime derived from SNe Ibn observations \citep[\eg][]{2022ApJMaeda}, especially when the ejected material is confined within a torus-like structure, as indicated by the solid line in Fig.\,\ref{fig:density profile from all binaries}. Such diverse CSM structures can have a significant impact on the resulting bolometric light curves and radio emission, providing a potential link between the observed SN properties and the prior binary evolution.

\subsection{Interaction-powered bolometric light curves}
\label{sec:Interaction-powered luminosity}

\begin{figure}
\begin{center}
\includegraphics[width=0.5\textwidth]{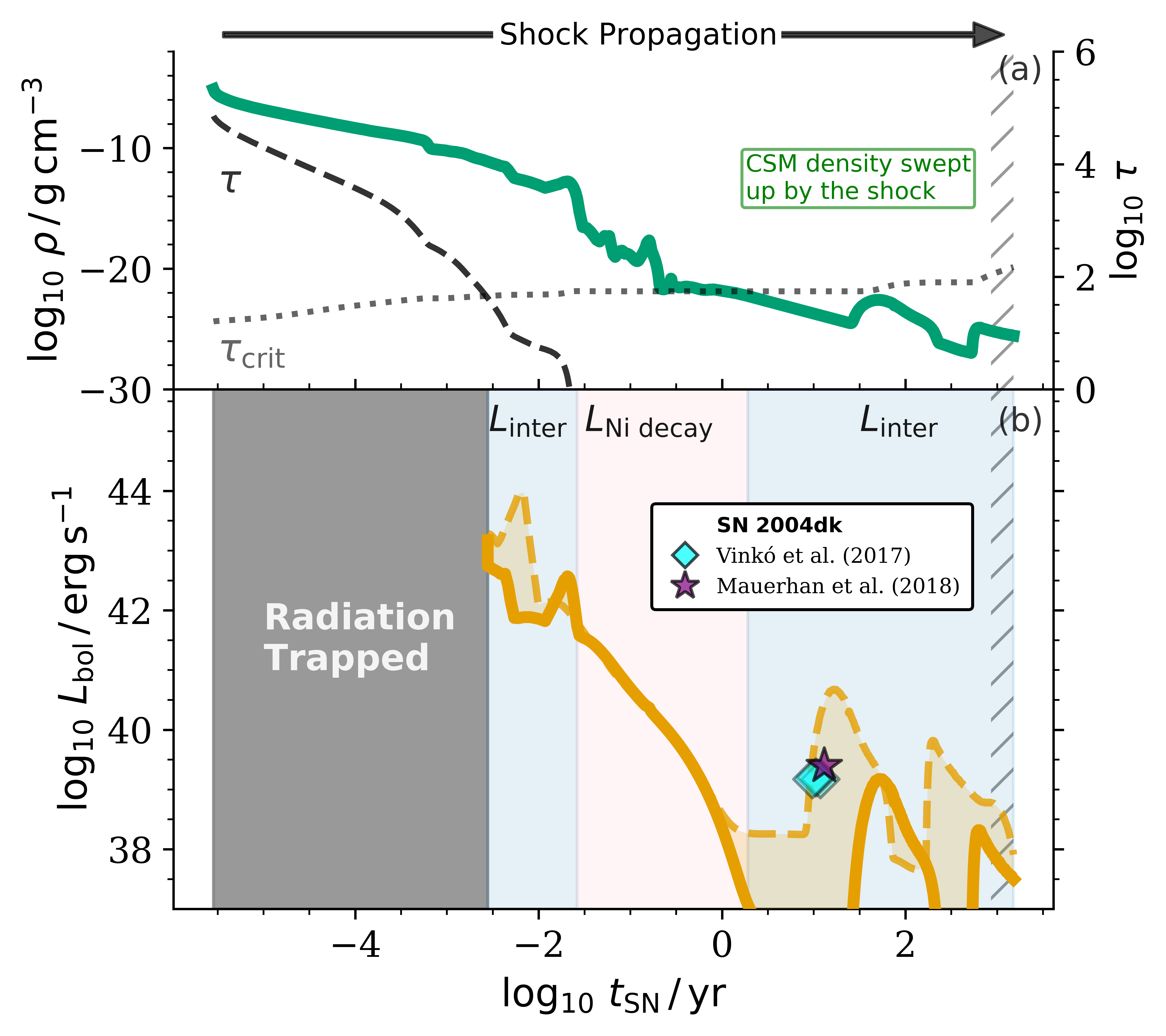}
\caption{Circumstellar density encountered by the shock and the corresponding bolometric light curve following the core collapse of an initially $2.75\,M_{\odot}$ helium star orbiting a NS companion. In panel (a), the green curve represents the density of the material being swept up by the shock as it propagates through the circumstellar medium, while the black dashed and grey dotted lines indicate the optical depth ($\tau$) ahead of the shock and its critical value ($\tau_{\rm crit}$), respectively. In panel (b), the orange solid and dashed curves represent the simulated bolometric light curves corresponding to the SN explosion energy of $10^{50}\,\rm erg$ and $10^{51}\,\rm erg$, respectively. The grey-shaded region marks the radiation-trapped phase where $\tau>\tau_{\rm crit}$, whereas the blue and pink regions indicate phases dominated by the interaction-powered and $^{56}{\rm Ni}$ radioactive decay luminosity, respectively. Optical observations of SN 2004dk from different works are shown as a purple star and cyan squares, respectively. The grey hatched region indicates the radial range where the CSM may be affected by the surrounding interstellar medium, and where the corresponding light-curve predictions become uncertain.}
\label{fig:interaction luminosity}
   \end{center}
\end{figure}

\begin{figure*}
\begin{center}
\includegraphics[width=1\textwidth]{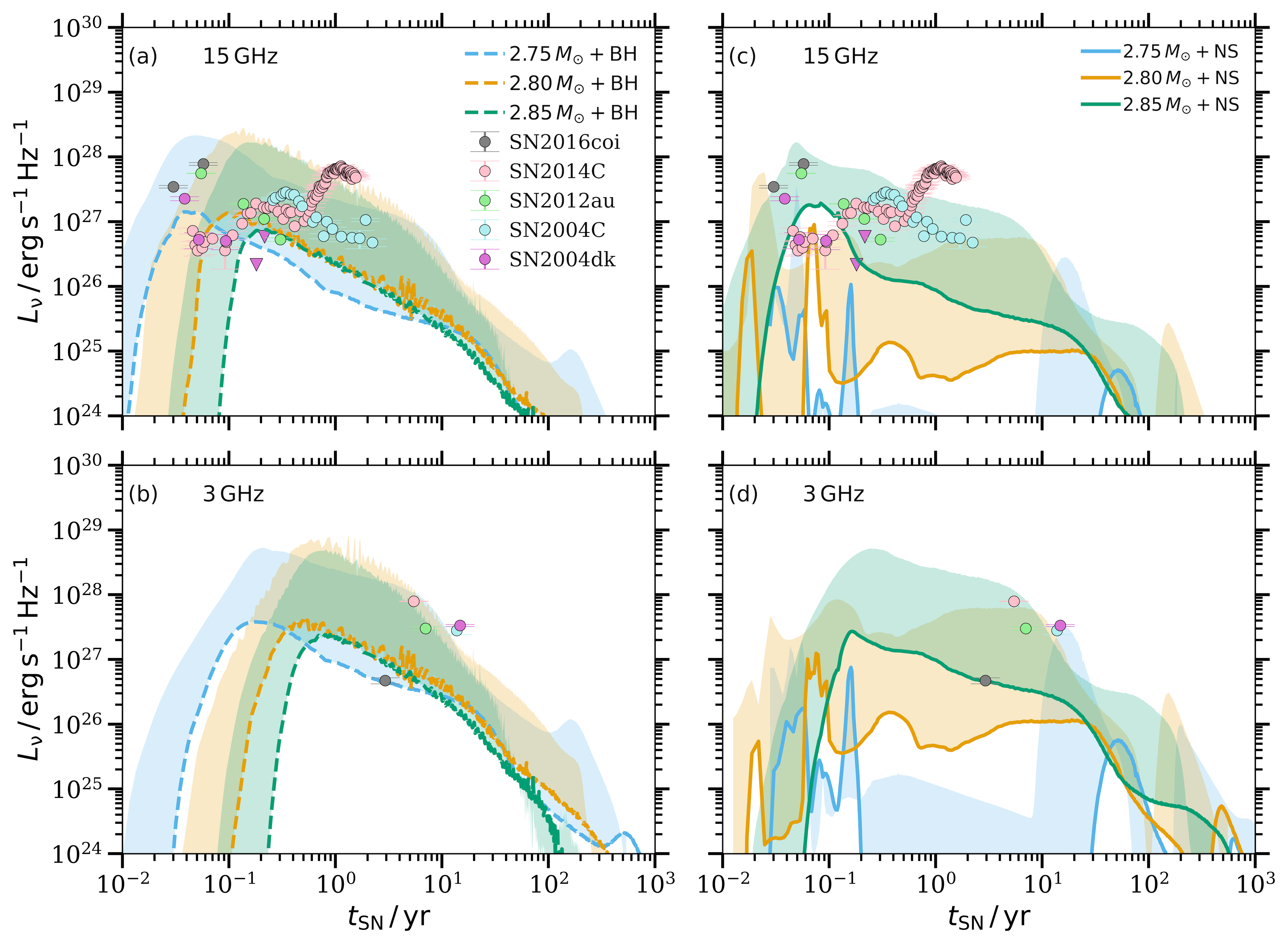}
\caption{Radio light curves at $15\,\rm GHz$ (upper panels) and $3\,\rm GHz$ (bottom panels) for distinct binary models. Solid and dashed lines represent the binary models with NS and BH companions, respectively. Binary models with helium-star masses of $2.75\,M_{\odot}$, $2.80\,M_{\odot}$, and $2.85\,M_{\odot}$ are shown in blue, orange and green colors, respectively. The shaded regions indicate the luminosity range spanned by explosion energies of $E_{\rm SN} = 10^{50}\,\rm erg$ (fiducial) and $10^{51}\,\rm erg$. Observed radio data are plotted as colored symbols, with error bars and downward triangles indicating upper limits. The observational data are taken from \citet{2012ApJWellons, 2014ApJKamble, 2017MNRASAnderson, 2019ApJTerreran, 2021ApJStroh}, and \citet{2022ApJDeMarchi}.
}
\label{fig:radio comparisons}
   \end{center}
\end{figure*}

Diverse CSM structures can give rise to corresponding variations in the light curves powered by the interaction between the SN ejecta and the CSM. For the fiducial binary model with a $2.75\,M_{\odot}$ helium star orbiting a NS companion, the SN ejecta mass is $M_{\rm ej} = 0.26 \,\rm M_\odot$, given the final stellar mass of $M_{1,f} = 1.66\,M_\odot$ for the stripped helium star (see Table\,\ref{tab:pre-after-MT}). The SN bolometric light curve is shown in Fig.\,\ref{fig:interaction luminosity}{\color{blue}b}, while Fig.\,\ref{fig:interaction luminosity}{\color{blue}a} shows the density of the CSM being swept up by the shock as it propagates through the isotropically symmetric CSM density profile established at core collapse, which is assumed to remain unchanged during the shock evolution. The optical depth of the CSM, $\tau = \int_{r}^{\infty} \kappa\,\rho_{\rm csm}\,\mathrm{d}r$, is plotted as a dark dashed line in Fig.\,\ref{fig:interaction luminosity}{\color{blue}a}. Here, we adopt electron-scattering opacity $\kappa = 0.2\,(1+\rm X(H))$, with $\rm X(H) = 0$ appropriate for helium rich material. In this model, the inner CSM is sufficiently dense to remain optically thick, with an optical depth $\tau > \tau_{\rm crit}$, as indicated by the grey-shaded region in Fig.\,\ref{fig:interaction luminosity}{\color{blue}b}. The critical value, $\tau_{\rm crit} = c\,/\,v_{\rm sh}$, corresponds to the condition for shock breakout, which takes place when $\tau = \tau_{\rm crit}$ \citep{2018PhRvDMurase}. 

Radiation trapped in this optically thick region is assumed to be released instantaneously at the moment of shock breakout, with the breakout luminosity estimated as $L_{\rm br} = \frac{1}{t_{\rm{br}} - t_{\rm i}} \int_{t_{\rm{i}}}^{t_{\rm br} } L_{\rm inter}\,\mathrm{d}t$. Here, $L_{\rm br}$ is the breakout luminosity at time $t_{\rm br}$. In the fiducial case with $E_{\rm SN} = 10^{50}\rm \,erg$, the breakout luminosity reaches $\sim 10^{43}\,\rm erg\, s^{-1}$ about one day after the SN explosion. Within the first $\sim 10\,\rm days$ after breakout, the light curve is dominated by the interaction-powered component (Fig.\,\ref{fig:interaction luminosity}{\color{blue}b}), showing features that closely trace the CSM density structure in Fig.\,\ref{fig:interaction luminosity}{\color{blue}a}. For instance, the sharp decline in the luminosity occuring at $\sim 10\,\rm days$ following the breakout results from a sudden drop in the CSM density. Subsequently, the light curve transitions to being powered primarily by the radioactive decay of $^{56}\rm Ni$. After about $1000\,\rm days$ ($\sim 3\,\rm yr$), the emission again becomes dominated by the interaction with the outer CSM. Two prominent late-time luminosity peaks at approximately $60\,\rm yr$ and $\sim 500\,\rm yr$ after the SN explosion correspond to density enhancements in the CSM (\cf Fig.\,\ref{fig:interaction luminosity}{\color{blue}a}). The grey hatched region in Fig.\,\ref{fig:interaction luminosity} marks epochs later than $\rm 1000\,yr$ after the explosion, when the shock reaches the material located at $\rm \,R > 10^{19}\,\rm cm$. Since the CSM at these distances may have been modified by the surrounding interstellar medium, the corresponding luminosity predictions become uncertain.

For comparison, a bolometric light curve with a higher SN explosion energy of $10^{51}\rm erg$ is calculated, as shown by the orange dashed line in Fig.\,\ref{fig:interaction luminosity}{\color{blue}b}. In this more energetic model, the shock propagates faster, resulting in generally higher luminosities and causing the subsequent interaction-powered peaks to occur at earlier phases. These late-time, interaction-powered rises in the modelled light curves provide a plausible explanation for the optical rebrightening observed in the SN 2004dk (as shown in Fig.\,\ref{fig:interaction luminosity}{\color{blue}b}), which was reported by \citet{2017ApJVinko} and \citet{2018MNRASMauerhan}. Although no signs of CSM interaction had been detected optically before the rebrightening, signatures of the CSM interaction were detected at other wavelengths, including radio emission (Fig.\,\ref{fig:radio comparisons}). 

Bolometric light curves for other binary models are shown in Fig.\,\ref{fig:bolometric_luminosity_appendixB} and exhibit a wide diversity, reflecting the diversity of their CSM structures. In these models, the interaction-powered luminosity might be comparable to the radioactive emission in the early phases, and it dominates the late-time emission in the light curves. Overall, the close correspondence between the light-curve morphology, the CSM density structure, and the progenitor’s radius evolution indicates that observed light curves can provide valuable diagnostics of the late-stage radius variations of massive stars and have the potential to constrain their mass-loss histories during the binary interactions.

\subsection{Interaction-powered radio light curves}

Radio light curves based on different binary models are shown in Fig.\,\ref{fig:radio comparisons}, where we assume an isotropically symmetric CSM structure. At a given frequency, the radio emission is strongly affected by the free-free absorption and synchrotron self-absorption, particularly in a dense CSM environment. The early-time rise of the radio peak is attributed to the decrease of the optical depth (or absorption) as the shock expands into a less dense CSM as shown in Fig\,\ref{fig:density profile from all binaries}. The relative shifts in the light-curve peaks among different models in each panel of Fig.\,\ref{fig:radio comparisons} can be attributed to the differences in CSM absorption from different models. For example, in Fig.\,\ref{fig:radio comparisons}{\color{blue}a}, the light-curve peak of the $2.75\,M_{\odot} + \rm BH$ binary model appears roughly $70\,\rm days$ earlier than in the $2.85\,M_{\odot} + \rm BH$ binary model. This difference results from the higher inner CSM density in the latter case (Fig.\,\ref{fig:density profile from all binaries}{\color{blue}a} and Fig.\,\ref{fig:density profile from all binaries}{\color{blue}c}), which leads to stronger absorption.

For a given CSM profile derived from the a specific binary model, the peak of the radio light curve appears earlier at higher frequencies. For instance, in the case of the $2.75\,M_{\odot} + \rm BH$ binary model, the radio emission at $15\,\rm GHz$ reaches its peak at around $\sim 14\,\rm days$ after core collapse (see blue dotted line in the panel of  Fig.\,\ref{fig:radio comparisons}{\color{blue}a}), whereas the $3\,\rm GHz$ emission reaches its maximum approximately about $\sim 50\,\rm days$ later (see blue dotted line in the panel of  Fig.\,\ref{fig:radio comparisons}{\color{blue}b}). This behaviour arises because high-frequency photons are more energetic, and can escape from the dense CSM at earlier times following the explosion. As the shock propagates outward, the surrounding CSM density gradually decreases, resulting in weaker absorption. Consequently, low-frequency photons are able to emerge at later times. In addition, for a fixed CSM environment, low-frequency emission exhibits higher peak luminosity due to the nonthermal spectrum (Eq.~\eqref{eqs.19}, see also \citealt{2002ARA&AWeiler}).

Compared to the models containing BH companions, systems with NS companions exhibit more complex radio light-curve morphologies (e.g. $2.75\,M_{\odot} + \rm NS$ and $2.80\,M_{\odot} + \rm NS$). These differences originate from the distinct CSM density structures surrounding the SN progenitors. As shown in the right panels of Fig.\,\ref{fig:density profile from all binaries}, the CSM profiles in Figs.\,\ref{fig:density profile from all binaries}{\color{blue}d} and {\color{blue}7e} display multiple density peaks, whereas other models show either a continuous distribution or a single detached density structure near the inner radius. Overall, the early-time rise in the radio emission is mainly driven by the decreasing absorption optical depths, whereas the late-time morphology is governed by the detailed CSM structure shaped by the previous binary interactions. Comparisons between the modelled radio light curves and observations of stripped-envelope SNe can therefore provide valuable constraints on the nature of the surrounding CSM and on the late-stage evolution of their progenitor systems. 

For comparison, we also investigate models with a higher explosion energy of $E_{\rm SN} = 10^{51}\,\rm erg$, in addition to the fiducial model with $E_{\rm SN} = 10^{50}\,\rm erg$. The shaded region indicates the range of radio emission spanned between the fiducial and high-energy explosion models. Overall, models with higher explosion energy exhibit stronger radio emission. In these energetic models, the shock propagates more rapidly and reaches lower-density regions of the CSM at earlier times. As a consequence, the early-time radio emission rises earlier due to the weaker absorption, and the subsequent re-brightening peaks also occur earlier than in the fiducial cases (e.g. $2.75\,M_{\odot} + \rm NS$ model in Figs.\,\ref{fig:density profile from all binaries}{\color{blue}c} and {\color{blue}7d}).

With peak luminosities reaching up to $\sim 10^{29}\,\rm erg\,s^{-1}\,Hz^{-1}$, our modelled radio light curves are comparable to the radio observations for some hydrogen-poor SNe. As indicated by the colored symbols in Fig.\,\ref{fig:radio comparisons}, the high-frequency radio emission (e.g. 15\,\rm GHz) is detected a few days after the SN explosion, corresponding to the early-time emission phase, while the lower-frequency emission (e.g. 3\,\rm GHz) tends to be more luminous at the later time and is associated with the late-time emission phase. Accounting for uncertainties in the SN explosion energy, the predicted radio luminosities span a broad range, and overlap with most of the observations at different frequencies (Fig.\,\ref{fig:radio comparisons}). For several events, such as SN 2016coi, SN 2012au, and SN 2004C, either the early- or late-time radio emission lies within the luminosity range predicted by plausible progenitor configurations and explosion energies. In the case of SN 2004dk, the $2.80\,M_{\odot} + \rm NS$ model with $E_{\rm SN} = 10^{51}\,\rm erg$ exhibits a radio evolution qualitatively similar to several key observational features, including an early decline followed by a later re-brightening phase.

A notable outcome of our modelling is that several binary progenitor models naturally produce multi-peaked radio light curves, for example the $2.75\,M_{\odot} + \rm NS$ and $2.80\,M_{\odot} + \rm NS$ models (Figs.\,\ref{fig:radio comparisons}{\color{blue}c} and {\color{blue}9d}). Such behaviour qualitatively resembles the complex radio variability observed in events like SN 2014C, which cannot be explained by smooth, single-peaked light curves. This suggests that SN ejecta interacting with a diverse binary-driven CSM may provide a promising approach for producing the observed diversity of radio light-curve morphologies. 

Overall, although our current models are not intended to quantitatively reproduce individual radio light curves, they demonstrate that SN–CSM interaction driven by binary mass loss may contribute to the diversity of radio behaviour observed in some stripped-envelope supernovae, including the presence of multi-peaked features. A broader exploration of binary configurations, particularly ultra-stripped progenitors with thin He-rich envelopes and more complex CSM geometries, will be crucial for more detailed modelling in the future.

\section{Discussion}
\label{Discussion}

In this work, we explore the diversity of SN light curves powered by the interaction between SN ejecta and the CSM formed through binary mass loss. Our models produce non-monotonic, multi-peaked light-curve structures, showing that binary-driven CSM can generate late re-brightening features qualitatively similar to those observed in some stripped-envelope SNe at optical wavelengths. In the radio band, our models give rise to a broad range of luminosities and complex, multi-peaked light-curve morphologies that resemble those seen in some interacting stripped-envelope supernovae. Our results suggest that structured CSM produced through binary evolution can contribute to the diversity of interaction-powered transients. They highlight a potential connection between the late-stage evolution of binary SN progenitors and the observable signatures of SN-CSM interaction.

The progenitor models in this work are considered as ultra-stripped supernova progenitors. However, only a handful of ultra-stripped SN candidates have been identified observationally to date, e.g. SN2005ek \citep{2013ApJDrout}, iPTF14gqr \citep{2018SciDe}, SN 2019dge \citep{2020ApJYao}, AT2019wxt \citep{2023ApJShivkumar}, SN2021agco \citep{2023ApJYan}, and SN 2023zaw \citep{2024ApJDas,2025ApJMoore}. Currently available observations do not yet provide strong constraints on the long-term interaction-powered signatures predicted by our models. Existing optical observations mainly probe the first few weeks after explosion and therefore primarily constrain material located close to the progenitor. While during this phase, the observed emission is generally dominated by radioactive heating, with some events also showing signatures of shock-cooling emission from extend material near the progenitor. In contrast, the interaction-powered signatures predicted by our models are expected to emerge predominantly at later times, typically years after the explosion in the optical band. In addition, radio observations of these ultra-stripped SN candidates remain very limited. In the few cases where radio follow-up has been reported (e.g. iPTF14gqr and AT2019wxt), the observations resulted in non-detections or upper limits rather than well-sampled radio light curves. We therefore compared our calculations with a broader sample of interacting stripped-envelope supernovae. The purpose of these comparisons is to illustrate that interaction between SN ejecta and structured CSM produced through binary evolution can generate a variety of light-curve morphologies qualitatively similar to those observed in some stripped-envelope supernovae. Future long-term multi-wavelength monitoring of ultra-stripped SN candidates, particularly at late times, will be crucial for probing structured CSM and constraining the long-term mass-loss history of stripped progenitors prior to core collapse.

Non-monotonic radio light curves have been reported in the models of \citet{2025ApJWu}, where such features arise for low-mass hydrogen-free progenitors only when a high outflow velocity of $1000\,\rm km\,s^{-1}$ is assumed. In contrast, the stripped helium stars in our binary systems are generally less massive and retain thinner helium-rich envelopes, resulting in more complex and diverse CSM density structures (see Sect.\,\ref{sec:Mirror effect}). Consequently, our models naturally result in diverse light-curve morphologies without requiring extreme outflow velocities.

In our models, the early-stage radio emission reaches luminosities comparable to those observed in some stripped-envelope SNe, while the early-time radio emission predicted by the models of \citet{2025ApJWu} tends to be fainter within the first few weeks after the explosion. This difference arises from the distinct CSM density profiles produced by different progenitor systems. In our models, most CSM density profiles exhibit a low-density inner cavity, formed during detached phases through low-density stellar winds. Furthermore, the RLOF outflow velocities obtained from our dynamical motions are typically a few times higher than the $100\,\rm km\,s^{-1}$ assumed in \citet{2025ApJWu}. As a result, our models produce overall lower CSM densities, leading to reduced synchrotron self-absorption and free-free absorption. Consequently, the radio emission becomes observable at earlier times and reaches its peak within the first few weeks after the explosion, with luminosities comparable to the early-time radio observations. The multiple sharp peaks appearing at early times in some of our models originate from highly time-variable mass loss shortly before core collapse. In our calculations, this variability emerges self-consistently from the binary evolution as a consequence of stellar radius variations, which produce substantial changes in the mass-transfer rate. Whether such fluctuations fully capture the complexity of pre-supernova mass loss in real systems remains uncertain, and additional processes, such as pulsational instabilities and wave-driven mass loss, may also contribute.

Dynamics of outflows from binary systems play a crucial role in shaping the CSM and determining the properties of the resulting observable transients. In this study, we model the outflow launched from the vicinity of the outer Lagrange point of the binary system under specific initial conditions that ensure the material becomes unbound and contributes to the CSM (see Sect.\,\ref{method:CSM}). However, in reality, different initial conditions of the outflowing material can lead to a variety of dynamical outcomes, including unbound outflows, fallback material forming a decretion disk, and self-intersecting trajectories that generate hydrodynamic shocks near the binary \citep{2019MNRASHubova}. Such complex outflow morphologies make it challenging to construct a realistic CSM structure. In addition, the density profiles produced by stellar winds and RLOF outflows are calculated separately in our approach, neglecting possible interactions between these two mass-loss mechanisms. Moreover, recent studies indicate that the wind mass-loss rates of stripped stars remain highly uncertain \citep{2016ApJTramper,2017MNRASYoon,2020MNRASSander,2023ApJGotberg,2025A&APauli}, which may affect the CSM density structure and, consequently, the observable properties of interacting SNe.

The velocity of the CSM is crucial in determining its density profile and, consequently, the resulting observational light curves. In this study, the CSM velocity is derived from the dynamics of the outflow launched from the outer Lagrange point of the binary system. This velocity is typically a few hundred $\rm km\,s^{-1}$ in our models (see Fig.\,\ref{fig:radial velocity of one particle}), which lies at the lower end of the range of CSM velocities inferred from observations of interacting SNe. In these events, the CSM is inferred to move at velocities of $\sim 10^2 - 10^3\,\rm km\,s^{-1}$, as indicated by the narrow emission lines observed in their spectra \citep{2008MNRASPastorello,2017ApJHosseinzadeh,2021ApJStrotjohann}. For binaries with shorter orbital periods, the velocity of the resulting CSM is expected to be higher. Additional acceleration processes occurring after the material is expelled from the system might further increase the CSM velocity. For instance, \citet{2024OJApTsuna} show that fast disk winds, driven by the Eddington-limited accretion onto the compact companion, can accelerate the CSM to velocities approaching $\sim 10^3\,\rm km\,s^{-1}$. In addition, after the SN explosion, radiation powered by the SN-CSM interaction may further accelerate the unshocked dense CSM ahead of the shock \citep{2023ApJTsuna}.

In this paper, we focus on the light curves calculated under the assumption of a spherically symmetric CSM with $\eta = 1$. However, outflows through the outer Lagrange point are expected to form a torus-like CSM structure \citep{2016MNRASPejcha, 2019MNRASHubova}. The resulting radio light curves vary with the viewing angle due to differences in the CSM density. From an edge-on viewing angle, the density of the torus-like CSM can be enhanced by a factor of $1/\eta$ compared to the spherical case, leading to stronger absorption and thus a delayed peak in the radio light curve, possibly accompanied by a slightly higher peak luminosity. In contrast, when the system is viewed face-on, the early radio rise originates from the interaction between the SN ejecta and the low-density stellar wind, whereas the subsequent emission is powered by the interaction with the dense torus-like CSM.

In addition to light-curve observations, the detection of narrow emission in the spectra provides further evidence for the presence of the CSM, which can be helium-rich (\eg SN 2000er; \citealt{2008MNRASPastorello}), or hydrogen-rich (\eg SN 2014C; \citealt{2017ApJMargutti}). In our model, with an initial helium-core mass of $2.74\,M_{\odot}$, a thin hydrogen-rich envelope of up to $0.1 \,M_{\odot}$ is retained. This hydrogen-rich material is stripped within the first $10\text{--}100$ years after the onset of mass transfer in the binary, and resides at distances beyond $10^{19}\,\rm cm$ by the time of core collapse (see Fig.\,\ref{fig:density profile from all binaries}), where the CSM may be modified by the interstellar medium. Therefore, the early SN–CSM interaction in our models is expected to be dominated by helium-rich CSM.  

However, our models might also exhibit hydrogen-line features at the early stage if the contribution from material that remains gravitationally bound after the CE ejection is taken into account. Three-dimensional magneto-hydrodynamic simulations of the CE interaction suggest that some hydrogen-rich materials can remain gravitationally bound to the inner binary following envelope ejection, potentially forming a circumbinary disk \citep{2022A&AMoreno, Wei2024A&A,2024A&AVetter,2025A&AVetter}. In post-CE binary systems, outflows produced during subsequent binary interaction could mix with the H-rich circumbinary material, particularly the disk lifetime can be comparable to the thermal timescale, or even nuclear timescale of the stripped helium star \citep{2023ApJTuna}. The resulting mixed CSM could therefore produce observable hydrogen-line features in the spectra. For binaries with a main-sequence accretor rather than a compact companion, an alternative explanation for the presence of hydrogen in the CSM may be mixing between helium-rich accreted material and the inflated hydrogen-rich envelope of the main-sequence star \citep{2025ApJWu}.

\section{Conclusion}
\label{Conculsion}

An increasing number of SNe exhibit clear signatures of interaction between the SN ejecta and dense CSM, offering valuable clues to the final evolutionary stages of their progenitors, although the physical origin of such CSM remains uncertain. In this work, we have focused on ultra-stripped progenitors and self-consistently modelled the CSM using the binary mass-loss history and the dynamical evolution of outflows launched through the outer Lagrange points. Based on the resulting CSM structures, we analytically predict the bolometric and radio light curves by following the shock dynamics arising from SN-CSM interaction. Our main findings can be summarised as follows.

\begin{itemize}
    \item The surface-radius evolution of ultra-stripped helium stars during the final evolutionary stage retains a clear imprint of core nuclear burning, with a delay corresponding to the thermal timescale. The surface of stars with thinner helium-rich envelopes responds more sensitively to internal nuclear activity, exhibiting stronger variability in their surface radius evolution. 
    
    \item Variations in the mass-transfer rate closely follow changes in the surface radius of stripped helium stars during their late evolutionary stages. Stripped helium stars with thinner envelopes exhibit more diverse mass-transfer histories.
    
    \item Up to $\sim 1\,\rm M_{\odot}$ of material may be expelled from the system due to the Eddington-limited accretion before core collapse, forming a dense CSM that subsequently interacts with the SN ejecta.
    
    \item The dynamics of outflows launched from the outer Lagrange points indicate that the binary tidal torque efficiently accelerates the escaping material, influencing the final CSM density distribution.
    
    \item Diversity of CSM density profiles (\eg detached CSM profiles or multi-peaked CSM density structures) reflects the changes in the total mass-loss rate, which in turn traces the progenitor's surface-radius evolution before core collapse. As a result, the CSM structure preserves key information about the nuclear activity and binary interaction history of the stripped stars.
    
    \item Our SN ejecta-CSM interaction models can produce the non-monotonic, multi-peaked light-curve morphologies that qualitatively resemble the optical re-brightening features observed in some stripped-envelope SNe (\eg SN 2004dk). 
    
    \item The models also yield early-time rising and multi-peaked radio light curves, qualitatively resembling some of the radio variability observed in stripped-envelope SNe. The early-time rise is mainly driven by the decreasing absorption optical depth as the shock expands into lower-density regions, whereas the late-time morphology depends on the detailed CSM structure shaped by the previous binary interactions. 
    
    \item At a given frequency, the radio emission peaks later in the models with denser CSM due to the stronger absorption, while for a specific CSM profile derived from a binary model, higher-frequency radio emission emerges earlier owing to the reduced absorption. 
    
\end{itemize}
We conclude that SN–CSM interaction in binaries hosting ultra-stripped progenitors can provide a possible channel for producing diverse light-curve morphologies. The resulting light curves reflect variations in the CSM structure shaped by binary-driven mass loss. Our work highlights a potential connection between the late-stage evolution of massive binary stars and the diversity of their explosive transients.

\begin{acknowledgements}
We thank the anonymous referee for the constructive comments that helped improve this manuscript. This work has received funding from the European Research Council (ERC) under the European Union’s Horizon 2020 research and innovation program (grant agreement No. 945806) and is supported by the Deutsche Forschungsgemeinschaft (DFG, German Research Foundation) under Germany’s Excellence Strategy EXC 2181/1–390900948 (the Heidelberg STRUCTURES Excellence Cluster). We acknowledge support from the DAAD-NINS Stellar Bridge program. Additional support was provided by the ISTA fellowship, as well as by the postdoctoral scholarship from the China Scholarship Council and the Deutscher Akademischer Austauschdienst (CSC–DAAD). EL acknowledges support through a start-up grant from the Internal Funds KU Leuven (STG/24/073) and through a Veni grant (VI.Veni.232.205) from the Netherlands Organization for Scientific Research (NWO).
\end{acknowledgements}

\bibliographystyle{aa}
\bibliography{./refs.bib}

\newpage
\clearpage

\begin{appendix}

\onecolumn

\clearpage
\section{CSM density structures of different models before collapse}
\label{Appendix: CSM density structures of different models in 2D}

\begin{figure}[!ht]
\centering

\begin{subfigure}{0.31\textwidth}
    \centering
    {\footnotesize $2.75\,M_{\odot}$ + BH}\\[1pt]
    \includegraphics[
        width=\linewidth
    ]{
        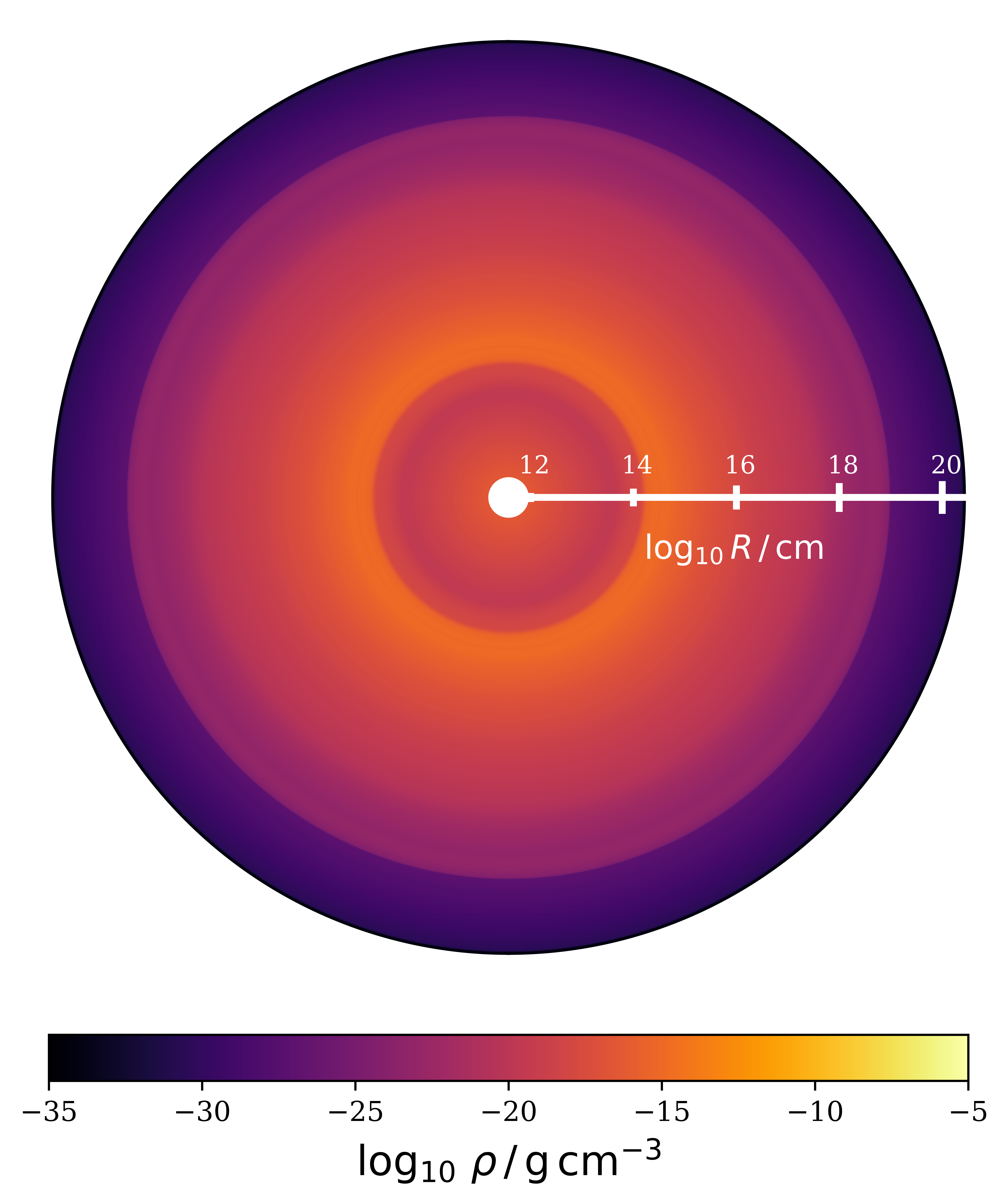
    }
\end{subfigure}
\hfill
\begin{subfigure}{0.31\textwidth}
    \centering
    {\footnotesize $2.80\,M_{\odot}$ + BH}\\[1pt]
    \includegraphics[
        width=\linewidth
    ]{
        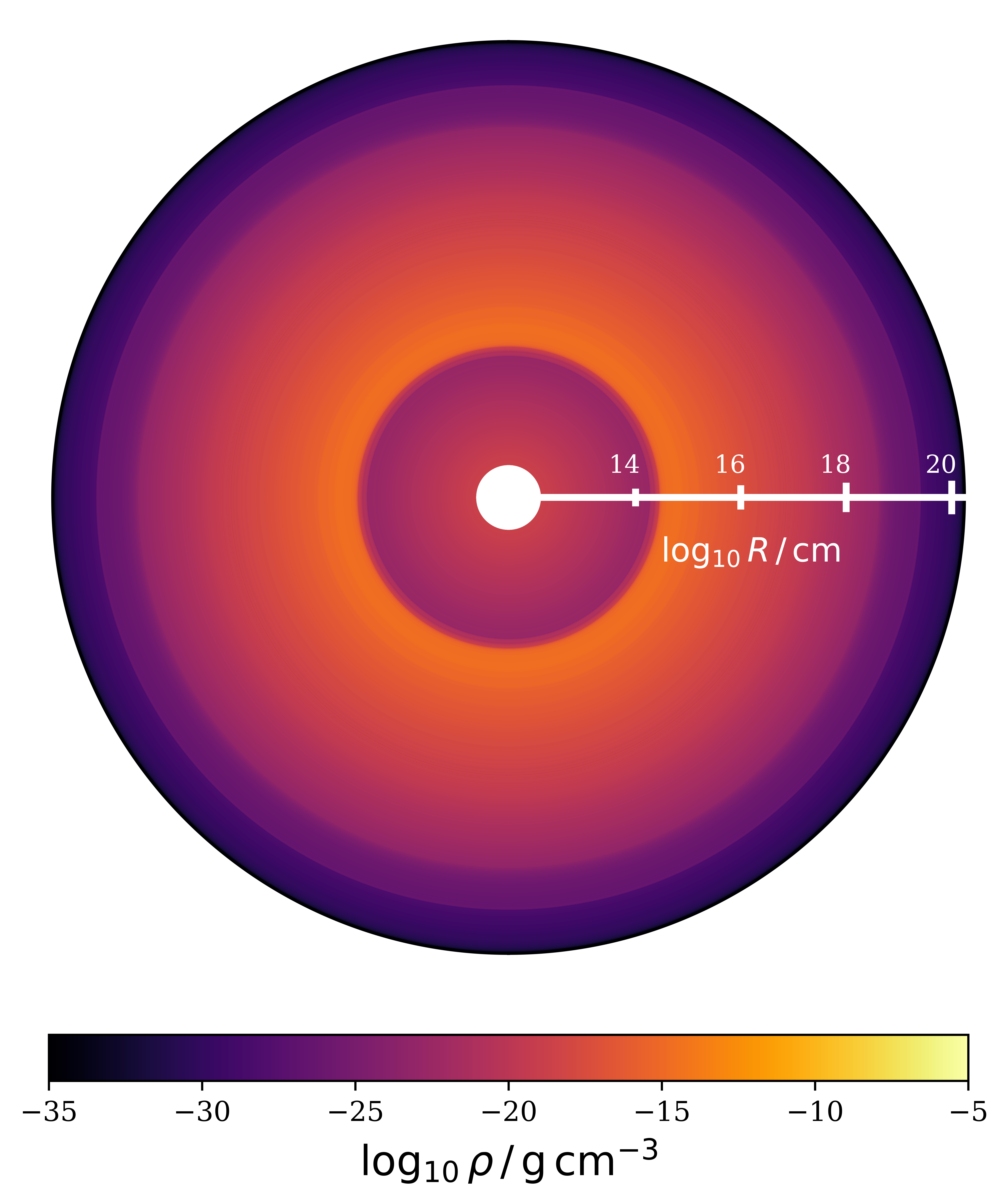
    }
\end{subfigure}
\hfill
\begin{subfigure}{0.31\textwidth}
    \centering
    {\footnotesize $2.85\,M_{\odot}$ + BH}\\[1pt]
    \includegraphics[
        width=\linewidth
    ]{
        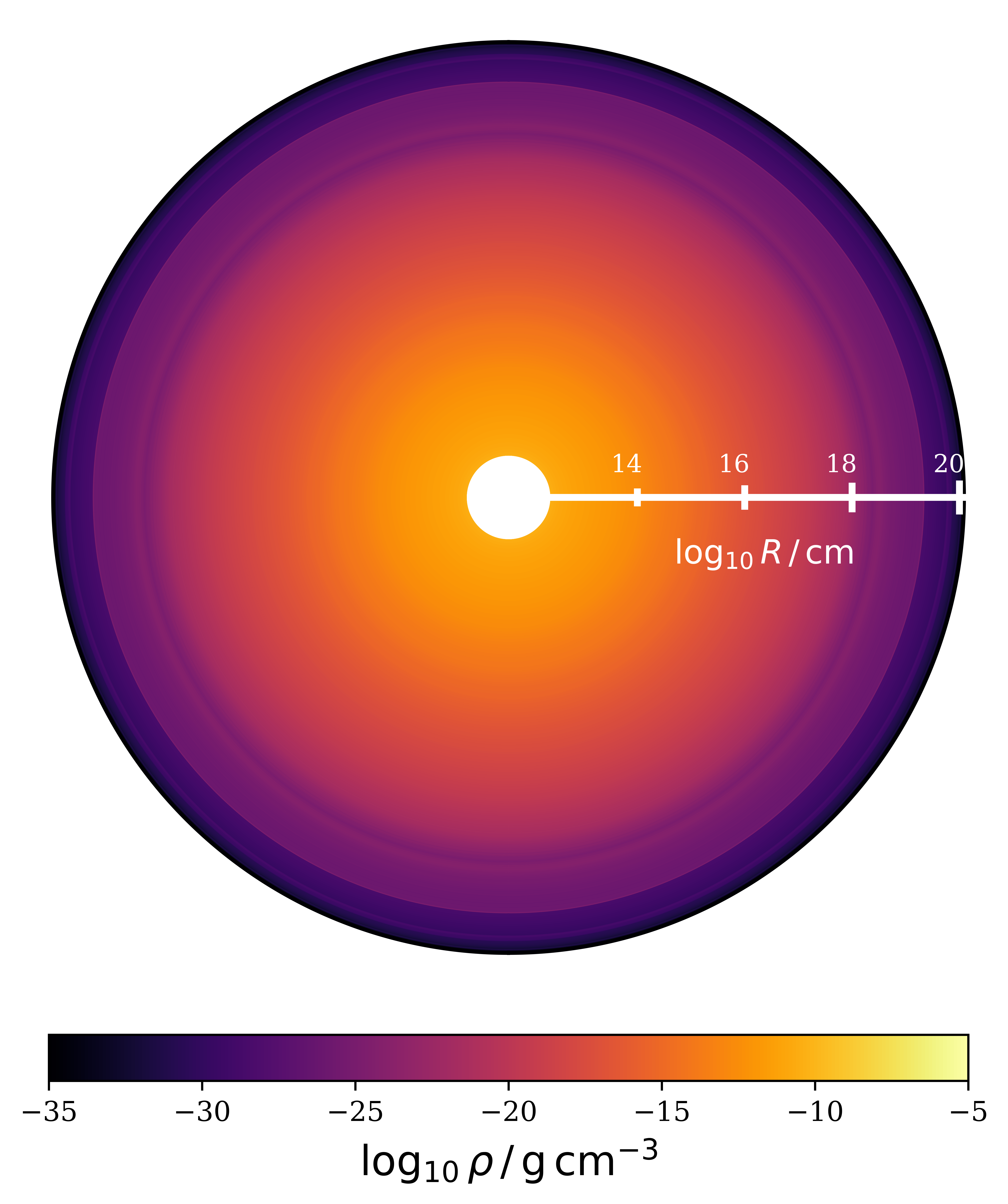
    }
\end{subfigure}

\vspace{0.4cm}

\begin{subfigure}{0.31\textwidth}
    \centering
    {\footnotesize $2.75\,M_{\odot}$ + NS}\\[1pt]
    \includegraphics[
        width=\linewidth
    ]{
        figs/Appendix/density_profiles/density_profile_circle_inferno_2_75_NS.png
    }
\end{subfigure}
\hfill
\begin{subfigure}{0.31\textwidth}
    \centering
    {\footnotesize $2.80\,M_{\odot}$ + NS}\\[1pt]
    \includegraphics[
        width=\linewidth
    ]{
        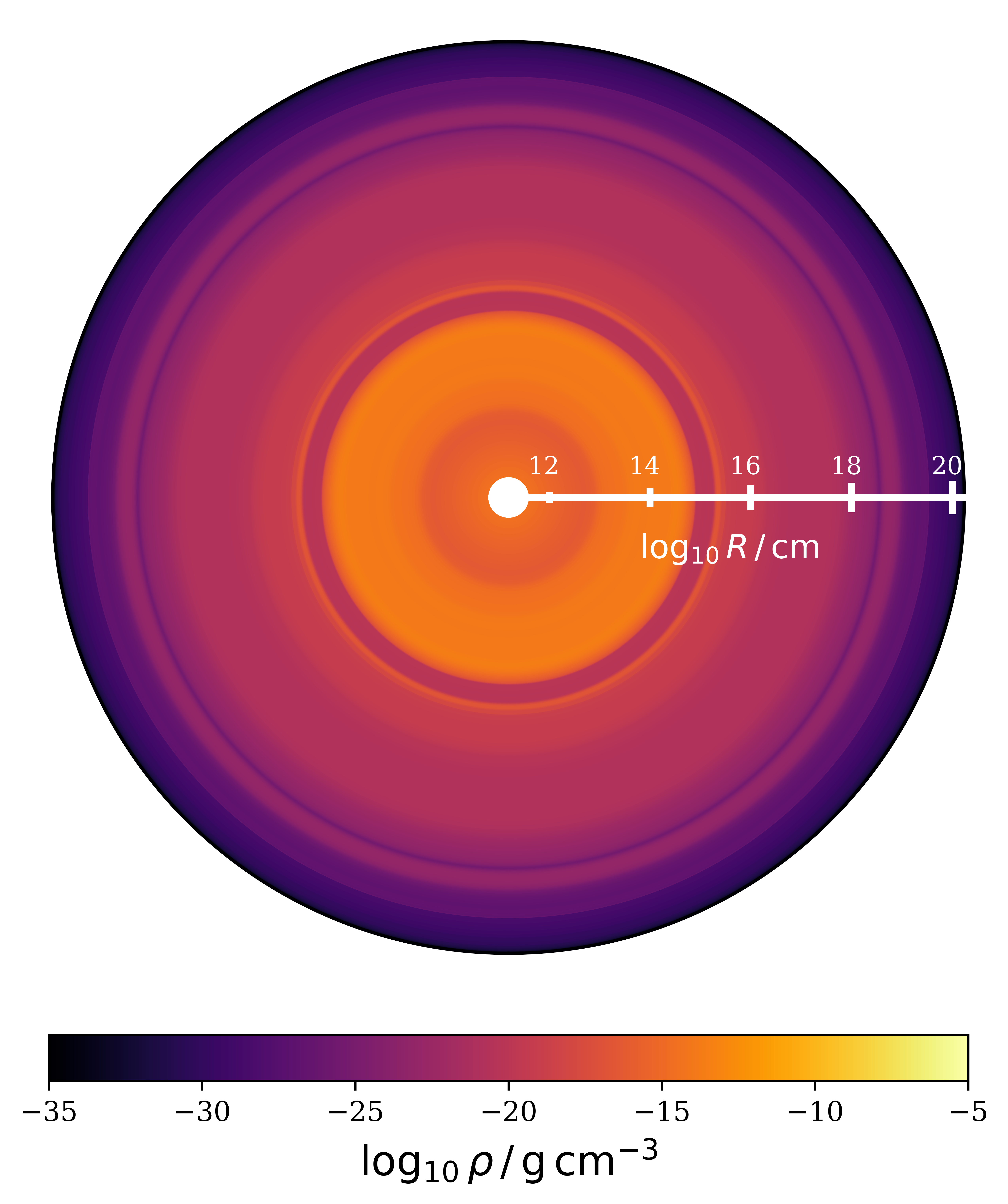
    }
\end{subfigure}
\hfill
\begin{subfigure}{0.31\textwidth}
    \centering
    {\footnotesize $2.85\,M_{\odot}$ + NS}\\[1pt]
    \includegraphics[
        width=\linewidth
    ]{
        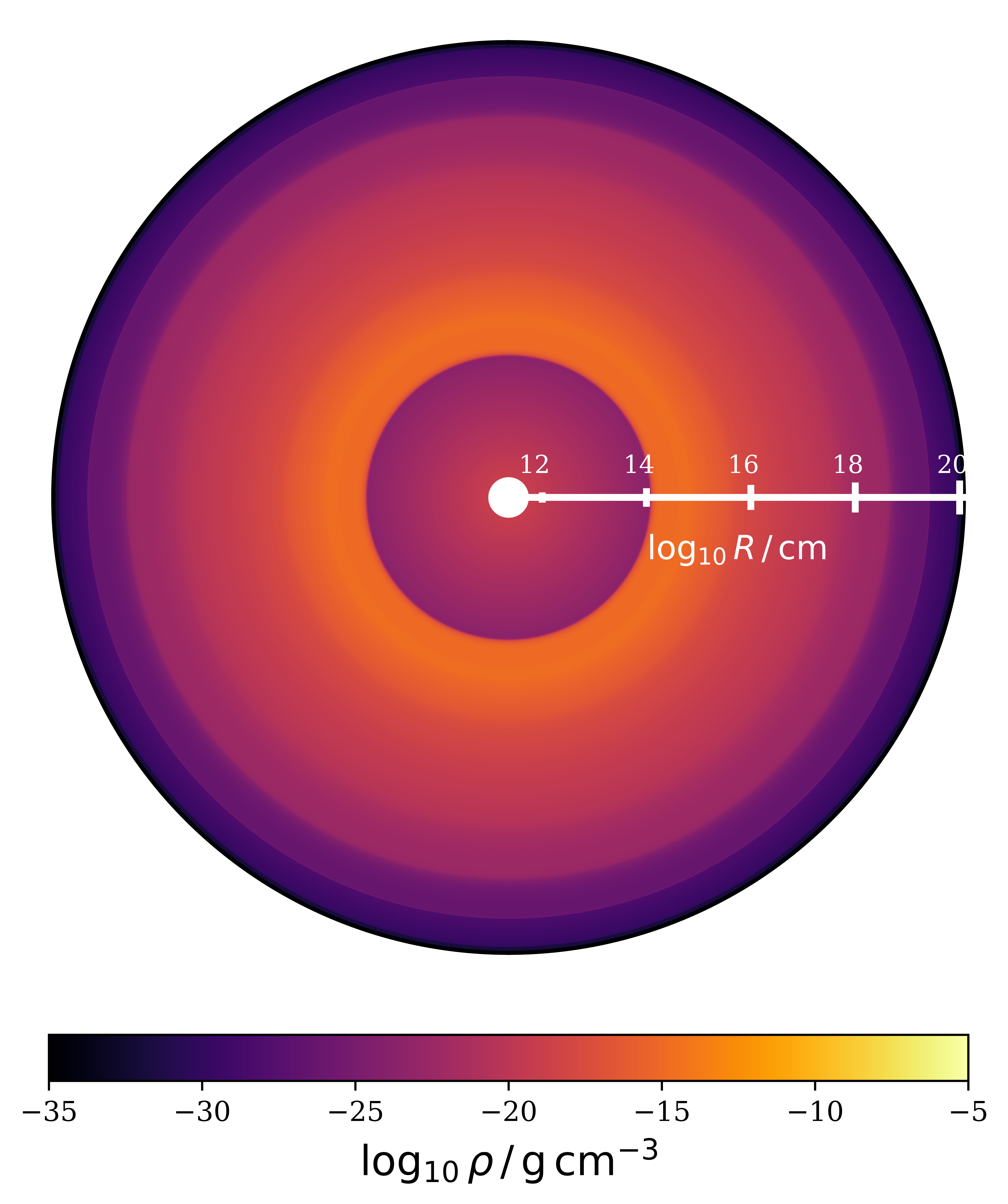
    }
\end{subfigure}

\caption{
Density distribution in the equatorial plane at core collapse
for the six binary models presented in this work, illustrating
the diversity of CSM structures arising from different binary
mass-loss histories. The CSM is assumed to be confined within
a torus-like geometry with $\eta = 0.1$.
}
\label{fig:density_structure}

\end{figure}

Two-dimensional CSM density maps for all binary models presented in this work are shown in Fig.\,\ref{fig:density_structure}, and the corresponding one-dimensional radial density profiles are shown in Fig.\,\ref{fig:density profile from all binaries}. These maps highlight the morphological diversity of the CSM
produced through different mass-loss histories, including continuous wind-like CSM structures, detached-shell CSM, and
non-monotonic multi-peaked structures. The structures at larger radii reflect earlier mass-transfer phases that occurred in the progenitor binary, while the inner regions preserve the signatures of the most recent episodes before core collapse (see Fig.\,\ref{fig:overall_density_2_75_NS} for more details).

\clearpage
\section{Bolometric luminosities of different binary models}
\label{Appendix: Bolometric luminosities of different binary models}

\begin{figure}[!ht]
\centering

\begin{subfigure}{0.31\textwidth}
    \centering
    {\footnotesize $2.75\,M_{\odot}$ + BH}\\[1pt]
    \includegraphics[
        width=\linewidth
    ]{
        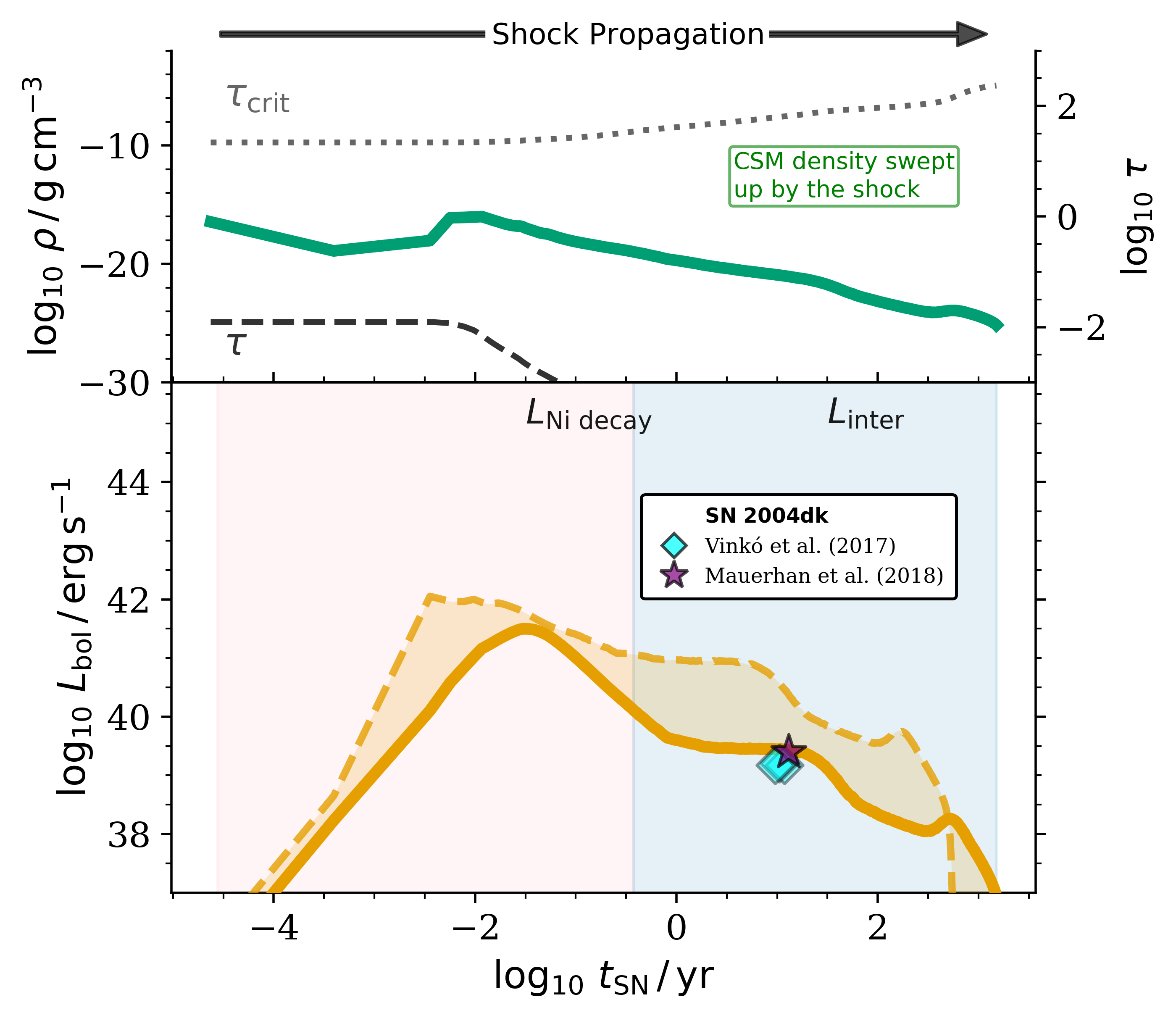
    }
\end{subfigure}
\hfill
\begin{subfigure}{0.31\textwidth}
    \centering
    {\footnotesize $2.80\,M_{\odot}$ + BH}\\[1pt]
    \includegraphics[
        width=\linewidth
    ]{
        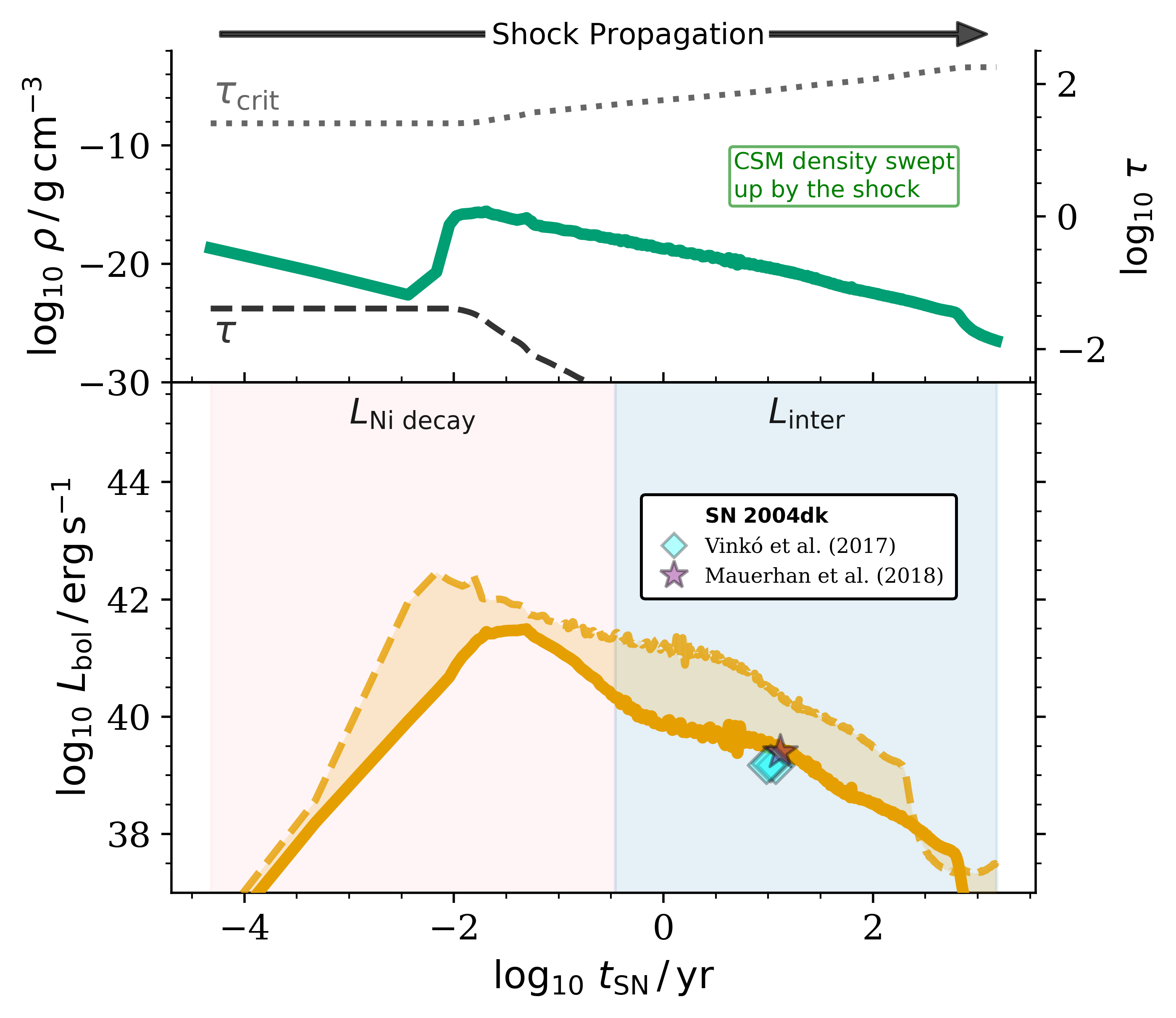
    }
\end{subfigure}
\hfill
\begin{subfigure}{0.31\textwidth}
    \centering
    {\footnotesize $2.85\,M_{\odot}$ + BH}\\[1pt]
    \includegraphics[
        width=\linewidth
    ]{
        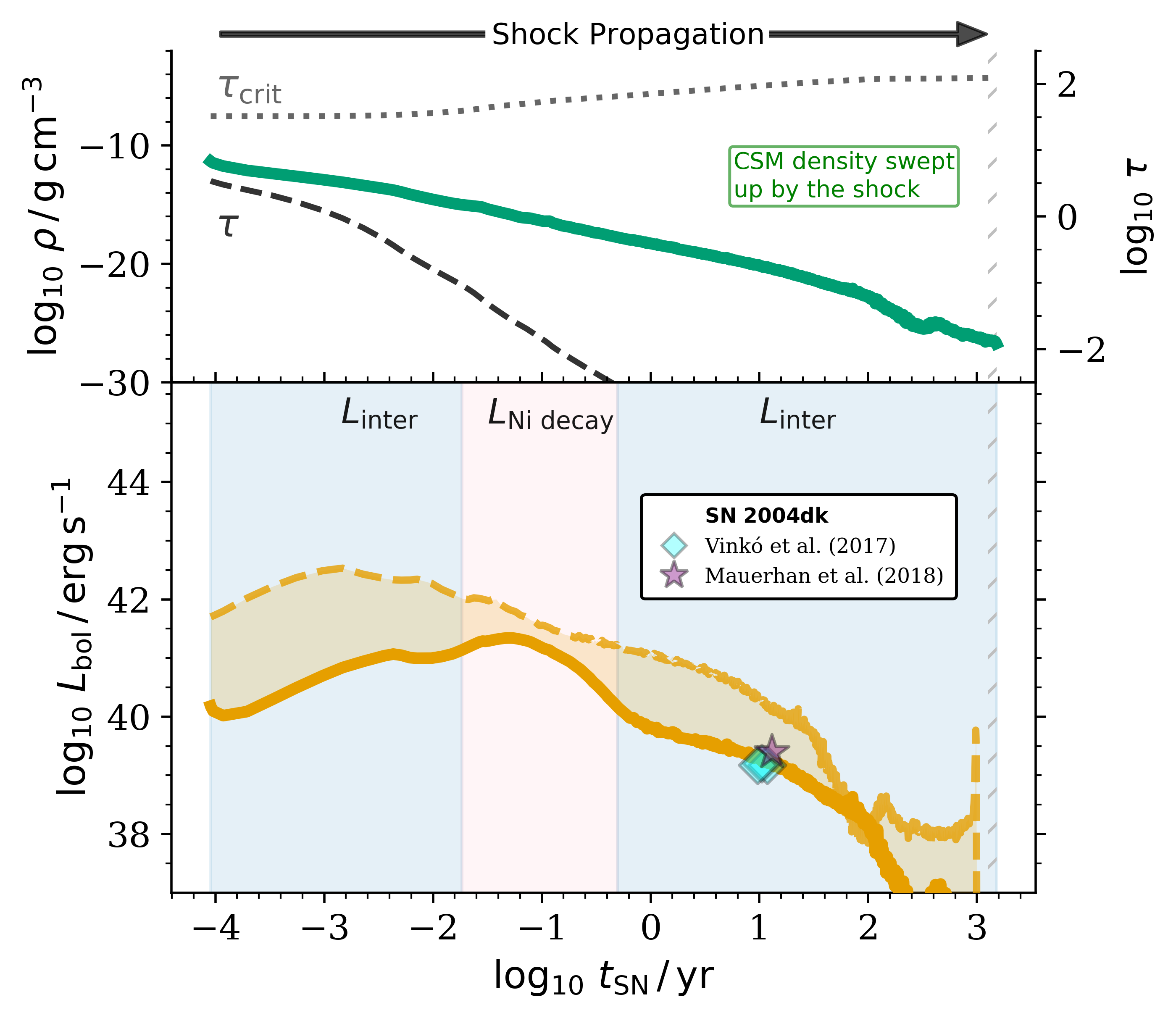
    }
\end{subfigure}

\vspace{0.4cm}

\begin{subfigure}{0.31\textwidth}
    \centering
    {\footnotesize $2.75\,M_{\odot}$ + NS}\\[1pt]
    \includegraphics[
        width=\linewidth
    ]{
        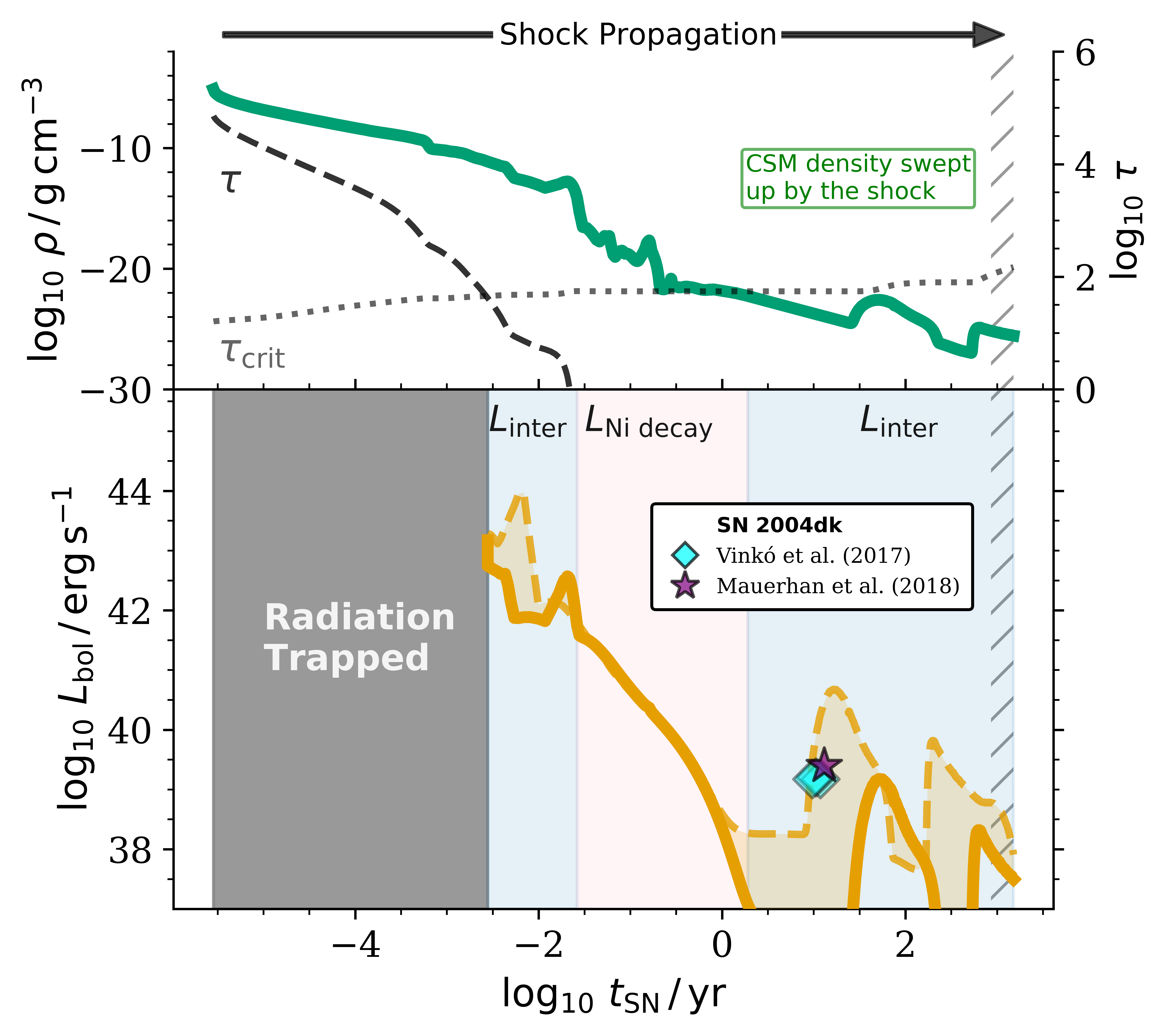
    }
\end{subfigure}
\hfill
\begin{subfigure}{0.31\textwidth}
    \centering
    {\footnotesize $2.80\,M_{\odot}$ + NS}\\[1pt]
    \includegraphics[
        width=\linewidth
    ]{
        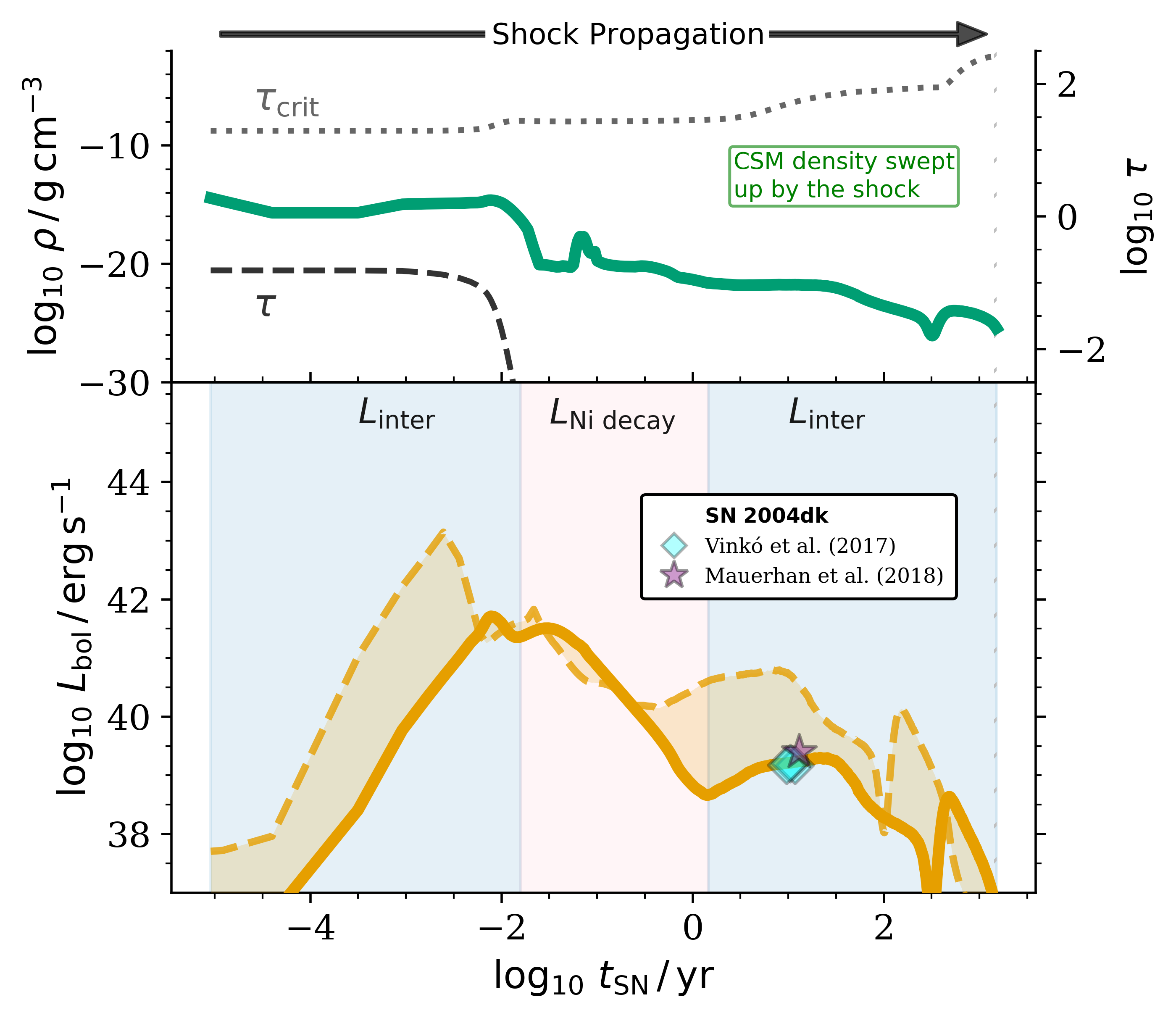
    }
\end{subfigure}
\hfill
\begin{subfigure}{0.31\textwidth}
    \centering
    {\footnotesize $2.85\,M_{\odot}$ + NS}\\[1pt]
    \includegraphics[
        width=\linewidth
    ]{
        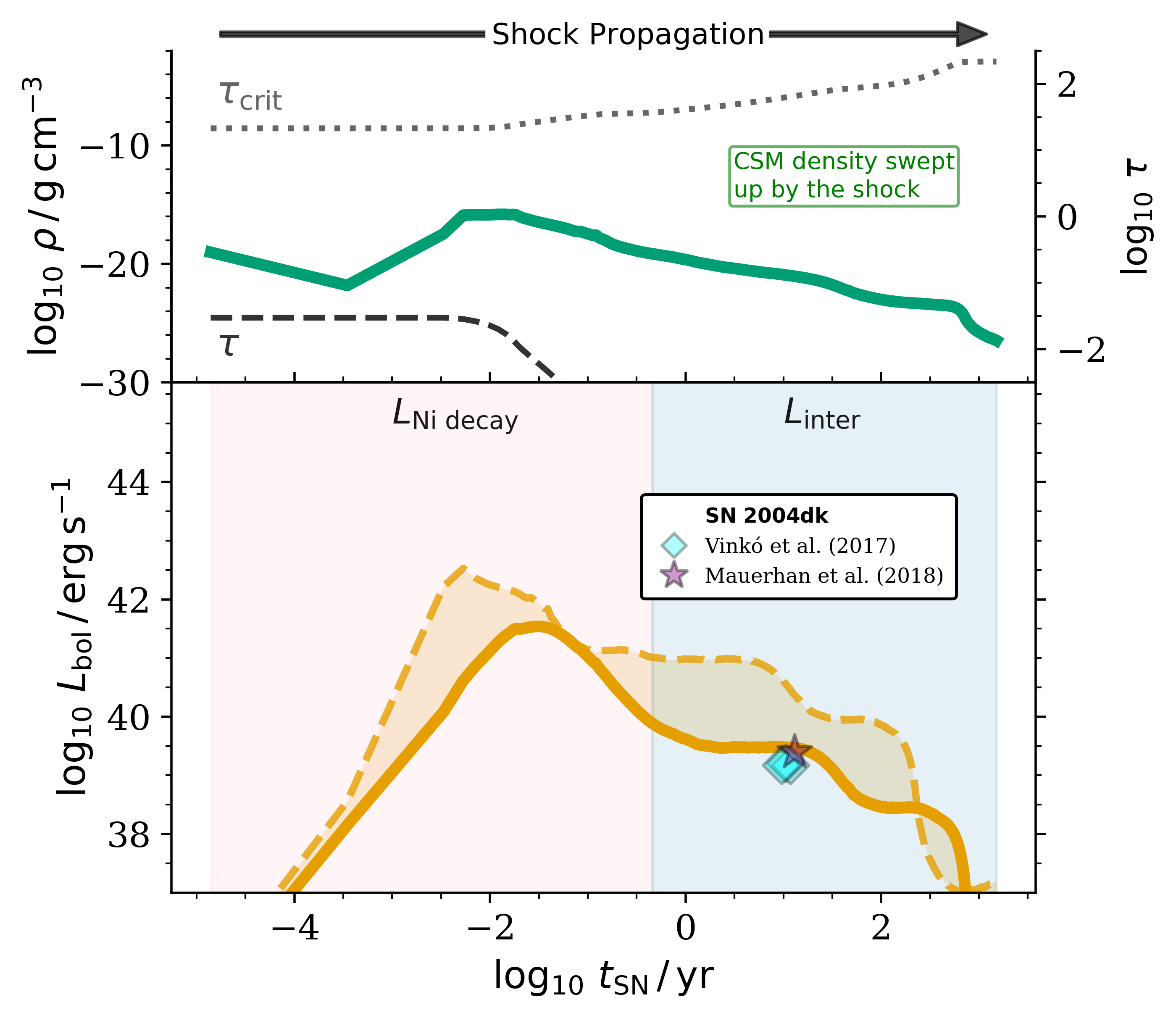
    }
\end{subfigure}

\caption{
Bolometric luminosities powered by the interaction between SN ejecta and the CSM produced in different binary models.
The orange-shaded region marks the phase in which the interaction-powered luminosity becomes comparable to that from radioactive $^{56}$Ni decay. All other colours and symbols are the same as in Fig.\,\ref{fig:interaction luminosity}.
}
\label{fig:bolometric_luminosity_appendixB}

\end{figure}

Bolometric light curves for all binary models are shown in Fig.\,\ref{fig:bolometric_luminosity_appendixB}. The
interaction-powered luminosity either dominates or is comparable to the radioactive component during the early phases, and it becomes the dominant contribution at late times. The late-time variations in the light curves reflect changes in the underlying CSM density profiles. Systems with stripped helium stars that retain thinner envelopes tend to produce interaction-powered light curves with non-monotonic, multi-peaked structures.

\end{appendix}

\end{document}